\documentclass[10pt,journal,compsoc]{IEEEtran}
\ifCLASSOPTIONcompsoc
  \usepackage[nocompress]{cite}
\else
  \usepackage{cite}
\fi

\ifCLASSINFOpdf
\else
\fi

\input{setup}
\usepackage[colorlinks,bookmarksopen,bookmarksnumbered,citecolor=red,urlcolor=red]{hyperref}
\usepackage{ragged2e}
\usepackage{enumitem,kantlipsum}
\usepackage{lettrine}
\usepackage{cite}
\usepackage{tablefootnote}
\usepackage[multiple]{footmisc}
\begin{document}

\title{Studying Ad Library Integration Strategies of Top Free-to-Download Apps}

\author{
Md Ahasanuzzaman,~Member,~IEEE,
Safwat Hassan,~Member,~IEEE, and
Ahmed~E.~Hassan,~Fellow,~IEEE
\IEEEcompsocitemizethanks{
\IEEEcompsocthanksitem Md Ahasanuzzaman, Safwat Hassan, and Ahmed E. Hassan are with the Software Analysis and Intelligence Lab
(SAIL), School of Computing, Queen's University, Canada. \protect\\
E-mail: md.ahasanuzzaman@queensu.ca,\protect\\ \{shassan, ahmed\}@cs.queensu.ca
}}

\IEEEtitleabstractindextext{%

\begin{abstract}
\justifying{In-app advertisements have become a major revenue source for app developers in the mobile app ecosystem. 
Ad libraries play an integral part in this ecosystem as app developers integrate these libraries into their apps to display ads.
In this paper, we study ad library integration strategies by analyzing 35,459 updates of 1,837 top free-to-download apps of the Google Play Store.
We observe that ad libraries (e.g., Google AdMob) are not always used for serving ads -- 22.5\% of the apps that integrate Google AdMob do not display ads. 
They instead depend on Google AdMob for analytical purposes.
Among the apps that display ads, we observe that 57.9\% of them integrate multiple ad libraries. 
We observe that such integration of multiple ad libraries occurs commonly in apps with a large number of downloads and ones in app categories with a high proportion of ad-displaying apps. 
We manually analyze a sample of apps and derive a set of rules to automatically identify four common strategies for integrating multiple ad libraries.
Our analysis of the apps across the identified strategies shows that app developers prefer to manage their own integrations instead of using off-the-shelf features of ad libraries for integrating multiple ad libraries. 
Our findings are valuable for ad library developers who wish to learn first hand about the challenges of integrating ad libraries.
}

\end{abstract}

\begin{IEEEkeywords}
Ad libraries, Integration strategies, Mining Android mobile apps, Google Play Store
\end{IEEEkeywords}}

\maketitle

\IEEEdisplaynontitleabstractindextext
\IEEEpeerreviewmaketitle

\section{Introduction}
\label{sec:intro}
\begin{figure}[!t]
	\centering
	\includegraphics[width=0.4\textwidth]{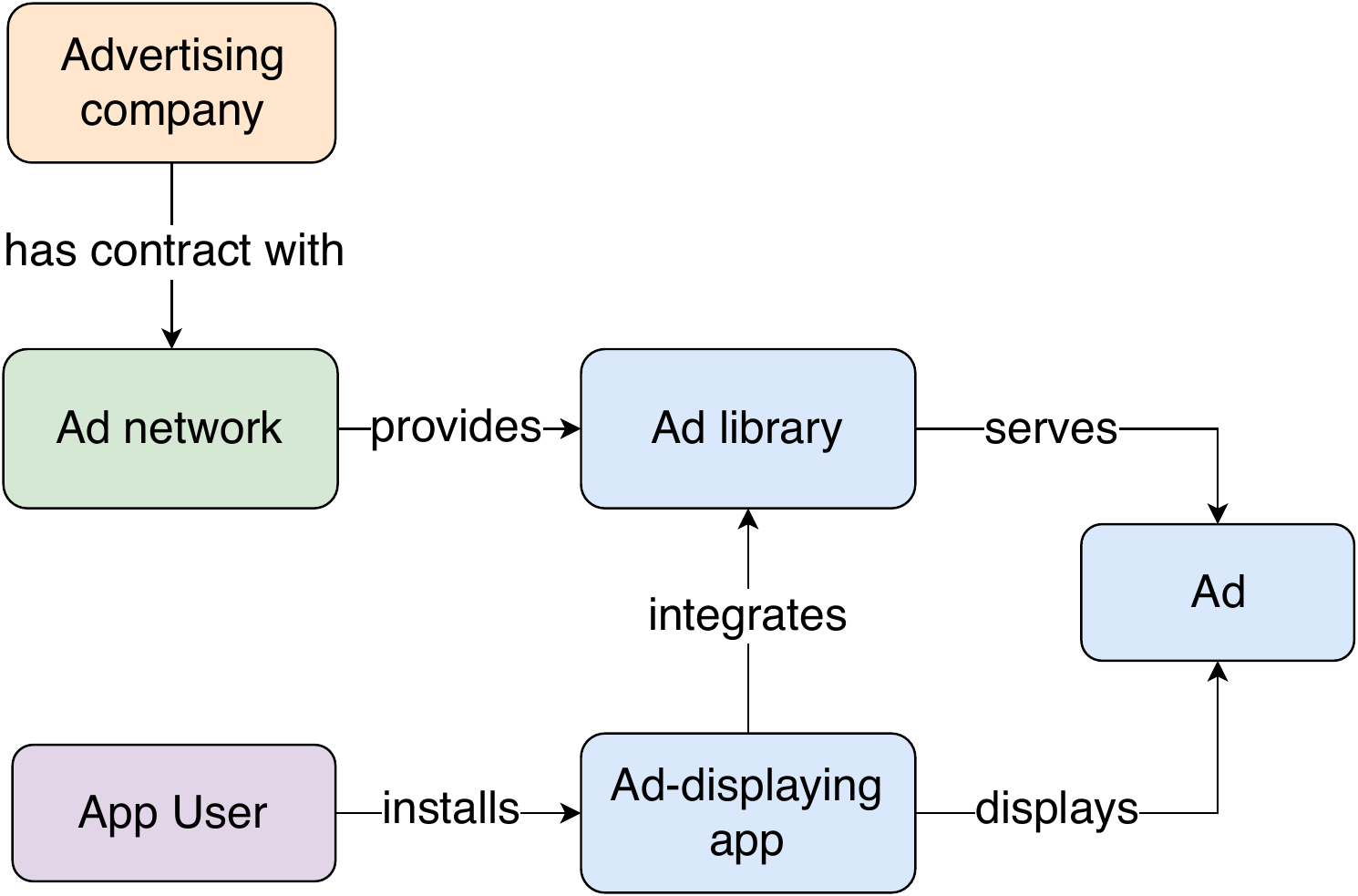}
	\caption{An overview of the in-app advertising model.}
	\label{fig:in_app_advertising_model}
\end{figure}

The mobile app market is continuously evolving  at a tremendous rate with billions of mobile app downloads every year~\cite{NumberOFDownloads_statista}.
The majority of the apps in app stores are free-to-download~\cite{app_stats}.
To earn revenue from these free-to-download apps, app developers primarily use an \textit{in-app advertising} model.
In this model, app developers display advertisements (\textit{ads}) to app users and earn revenue based on the number of displayed ads and the interactions of users with these ads.
The in-app advertising model is a growing market with a forecasted revenue of over \$200 billion by 2021~\cite{App_Annie_In_App_Growing_Market}.
Figure~\ref{fig:in_app_advertising_model} presents an overview of the in-app advertising model.
The in-app advertising model consists of four main components: 
(1) \textit{advertising companies} that pay for the display of ads for promoting their products, 
(2) \textit{ad-displaying apps} that display ads and earn revenue from the displayed ads, 
(3) \textit{mobile ad networks} which act as a bridge between advertising companies and ad-displaying apps, 
and (4) \textit{users} who use the ad-displaying apps and interact with the displayed ads.

To display ads, app developers need to register with an ad network (e.g., Facebook Audience Network) and integrate into their app an \textit{ad library} that is offered by the ad network.
The objectives of an \textit{ad library} is to manage the communication with an ad network, to display ads on a user's device, and to track user interactions with the displayed ads.

Since in-app advertising is a growing market, many ad networks are emerging in this market with their own ad libraries.
In this competitive market, app developers select an ad network that maximizes their revenue (e.g., offering a high fill rate\footnote{Fill rate is the ratio of the number of displayed ads over the number of requested ads.})~\cite{ad_library_selection}.
To earn more app revenue, app developers integrate multiple ad libraries with their apps to increase the fill rate~\cite{israel_app_rating_ads_library}.

Despite the integral role of ad libraries in the mobile app ecosystem, prior studies have not examined how these libraries are integrated into mobile apps and how app developers handle multiple ad libraries.
In this paper, we perform an in-depth study of the common strategies of integrating such ad libraries in the top free-to-download apps in the Google Play Store. 
Our study can help ad library developers understand the common challenges of integrating multiple ad libraries into mobile apps. 
Hence, ad library developers can improve the design and the offered features of their ad libraries to ease the ad library integration process.

To study the integration strategies for ad libraries, we analyzed 35,459 updates of 1,837 top free-to-download apps across all the categories of the Google Play Store.
Our scope for this research is to study how popular apps integrate ad libraries using standard practices.
In particular, we studied such strategies along with the following research questions (RQs): 

\noindent \textbf{\textbf{RQ1:} \textit{What are the characteristics of apps which integrate multiple ad libraries?}}

\noindent The integration of multiple ad libraries occurs commonly in the apps with a large number of downloads and ones in app categories where a high proportion of apps integrate ad libraries.

\noindent \textbf{RQ2:} \textbf{\textit{How do app developers integrate multiple ad libraries with their apps?}}

\noindent We manually examined a statistically representative random sample of ad-displaying apps (62) that integrate multiple ad libraries and derived a set of rules to automatically identify (four) strategies that app developers employ for integrating multiple ad libraries:~\textit{(1) external-mediation strategy} (app developers use an external-ad-mediator package that is provided by an ad library and do not write their own code to integrate other ad libraries),~\textit{(2) self-mediation strategy} (app developers write their own centralized code (self-mediator) to integrate ad libraries),~\textit{(3) scattered strategy} (app developers scatter their code across the different app screens), and~\textit{(4) mixed strategy} (app developers use both the external-mediation strategy and the scattered strategy).

We document the definition, example app, the benefits, and drawbacks of each identified strategy for integrating multiple ad libraries.
Developers of ad libraries can leverage our strategies to ensure that their ad libraries can support the varying needs of ad-displaying apps.

\vspace{0.1cm}
\noindent \textbf{Paper organization.}
\noindent Section~\ref{sec:Methodology} describes our data collection process. 
Section~\ref{sec:Data_Characteristics} discusses the characteristics of our dataset. 
Section~\ref{sec:Results} presents the results of our study.
Section~\ref{sec:Discussions} discusses how app developers maintain their integrated ad libraries over time.
Section~\ref{sec:Implications_And_Discussions} describes the implications of our work. Section~\ref{sec:Limitations_And_Threats} describes threats to the validity of our observations. Section~\ref{sec:Related_Work} discusses related work, and Section~\ref{sec:Conclusion} concludes the paper.

\section{Data collection}
\label{sec:Methodology}
\begin{figure}[!t]
    \centering
    \includegraphics[width=0.45\textwidth]{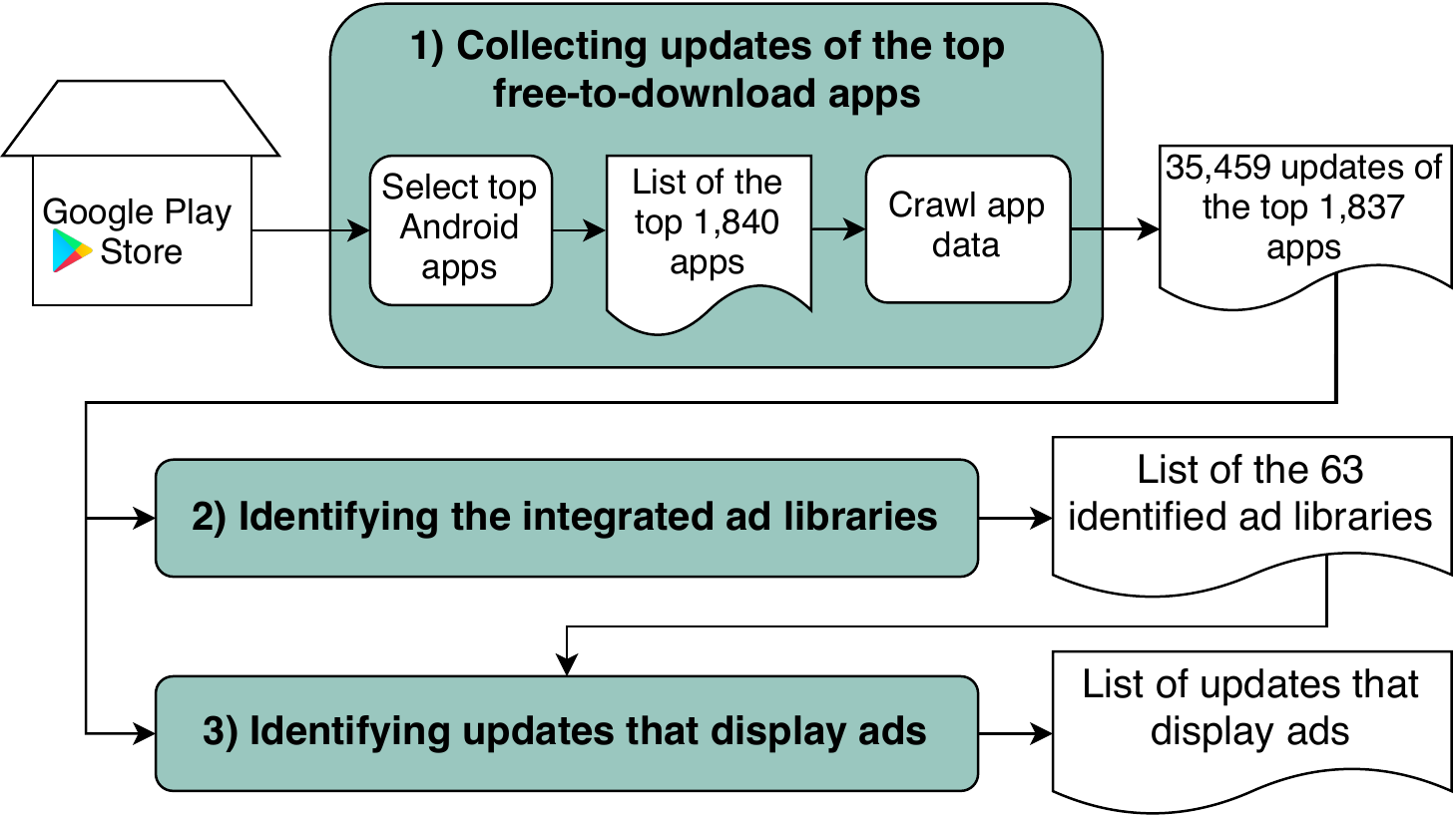}
    \caption{An overview of our data collection process.}
    \label{fig:methodology}
\end{figure}

This section describes our process for collecting ad library data.
Figure~\ref{fig:methodology} represents an overview of our data collection process.
As shown in Figure~\ref{fig:methodology}, we first collected the updates of top free-to-download apps in the Google Play Store.
Then, we identified ad libraries that are integrated by the apps in these updates.
Finally, we identified the updates that display ads.
We briefly highlight each step below.

\subsection{Collecting updates of the top free-to-download apps}

\textbf{Step 1: Select top Android apps.} 
In our study, we focus on the top free-to-download apps as these apps have a large user-base.
Hence, these apps are likely to follow the in-app advertising model to earn revenue.
Moreover, these apps are likely to carefully maintain their ad integration code in order to ensure that they do not lose any ad revenue.
To obtain the list of popular apps, we used the App Annie's report~\cite{AppAnnie_App_Selection} that lists the popular apps across the 28 categories (e.g., Games) in the Google Play Store in 2016. 
Then, we selected the top 100 apps in each app category so that our study does not have any bias due to variances across the different app categories.
During the app selection process, we found that 746 apps were already removed from the Google Play Store at the start of our study period and 214 apps were repeated across the app categories. 
In total, we selected 1,840 apps and downloaded all their deployed updates for our study.

\noindent\textbf{Step 2: Crawl app data.}
We ran a custom crawler (based on the Akdeniz’~\cite{crawler} Google Play crawler) for 18 months from April 20$^{th}$ 2016 to September 20$^{th}$ 2017 to collect the deployed updates of our studied apps. 
To study any changes (e.g., code changes) of an app, we need at least two updates of the app. We observed that three apps have only one update during our study period. Therefore, we removed these three apps from our study. 
Finally, we collected 35,459 updates of the 1,837 top free-to-download apps.

\subsection{Identifying the integrated ad libraries}
App developers integrate many third-party libraries and identifying an ad library package from these third-party libraries is a non-trivial task.
To identify an ad library package, we followed a similar approach to the exhaustive one that is employed by Ruiz et al.~\cite{updates_ad_library}. We detail our process below.

First, we converted the APKs of the collected updates to JARs using the dex2jar tool~\cite{Dex2jar}.
Then, we used the BCEL tool~\cite{BCEL} to extract the fully qualified class names (i.e., the class name and the package name) of all classes in the generated JARs. 
Since prior studies show that an ad library's packages or class names contain the term ``ad'' or ``Ad''~\cite{Li_Common_Library}, we filtered the fully qualified class names using the regular expression ``[aA][dD]''. 
However, this exhaustive regular expression matches many class names that are not related to ad libraries (e.g., \code{com.fbox.load.ImageLoad}).
Hence, to identify ad libraries, we followed Ruiz et al.'s~\cite{updates_ad_library} approach by manually verifying online the package name of each of the matched classes.
We manually verified 303 packages on the web. 
In total, we identified 63 ad libraries. 
\subsection{Identifying updates that display ads}
In the previous step, we identified the list of the integrated ad libraries. 
However, integrating an ad library in an update does not necessarily imply that the update displays ads (e.g., ad libraries can be used for analytical purposes as we discovered in our study). 
To identify the updates that display ads, first, we identified the app screens (as ads need to be displayed through the app screens).
Then, we identified the screens that display ads.
The details of our approach are as follows.

\noindent\textbf{Step 1: Identify app screens.}
To create a single app screen, app developers write the required functionality of the screen in a java class which is known as an \textit{Activity}. 
Then, app developers define the app screens (i.e, activities) in the \textit{AndroidManifest.xml} file using the~``$<$${activity}$$>$''~tag~\cite{android_manifest}.
Hence, to identify the app screens, we parsed the \textit{AndroidManifest.xml} file and listed all the defined activities (using the~``$<$${activity}$$>$''~tag) and their corresponding classes.

\noindent\textbf{Step 2: Identify the screens that display ads.}
First, we identified the integrated libraries in every screen using the BCEL tool~\cite{BCEL}.
Then, we identified screens that display ads if the screen code invokes the display method in the integrated ad library (e.g., calling the showAd() method). 
To identify the display ad methods, we read the documentation of the studied ad libraries and summarized the list of methods that are used for displaying ads. 
In our replication package\footnote{https://github.com/SAILResearch/suppmaterial-18-ahsan-ads\_consumer\_apps}, we added a list of such methods for each studied ad library.
Finally, we flagged an update as an ad-displaying update if the update contains at least one screen that displays an ad. 

At the end of this step, we identified all updates that display ads.

\section{Data characteristics}
\label{sec:Data_Characteristics}
In this section, we describe the characteristics of our dataset in terms of (1) ad-displaying functionality, (2) app category, and (3) integrated ad libraries.

\noindent\textbf{Ad libraries are not only used for serving ads but also for analytical purposes.}
The studied apps can be classified into two main categories: (1) \textit{ad-displaying} apps (i.e., apps that integrate ad libraries to display ads) and (2) \textit{non-ad-displaying} apps (i.e., apps that do not display ads).
Table~\ref{tab:stats_app_serve_ads} describes our dataset.

 As shown in Table~\ref{tab:stats_app_serve_ads}, non-ad-displaying apps can be of two types: (1) apps that integrate ad libraries but do not display ads and (2) apps that do not integrate ad libraries. We observe that 22.5\% of the non-ad-displaying apps belong to type 1 (i.e., integrate ad libraries but do not display ads), and all of these apps integrate the Google AdMob ad library.  We also identified 77 apps (4.2\% of the studied apps), where we observe that the apps contain ad library packages, but the static analysis tool could not find any method call to ad library packages. Of these 77 apps, we observe that 69 apps use native code. Studying native apps using static analysis tools is difficult and could introduce false positive cases in our analysis.  Therefore, in this paper, we focus on studying the apps (1,076 apps) that our static analysis approach identifies a call to show-ad methods of ad libraries from app code.

In our further analysis of the non-ad-displaying apps that integrate ad libraries, we observed that all these apps integrate the Google AdMob ad library for analytical purposes. We observed that analytical libraries (e.g., Google Analytics and AppsFlyer analytics) were dependent on the Google AdMob ad library for uniquely identifying a user’s device. Table~\ref{tab:stats_third_party_libs_use_Ads_Sdk} shows the top ten used third-party libraries that depend on the Google AdMob ad library (for the studied 154 apps) to identify a user’s device. For example, the Google Analytics library depends on the package ``com.google.android.gms.ads.identifier''~\cite{ads_identifier_package} of the Google AdMob ad library which provides the functionality to generate an Android Advertising ID (AAID) to identify a user’s device instead of using a user’s personal information (e.g., IMEI number or device MAC address -- a practice that is not recommended by Google)~\cite{android_advertising_id,DBLP:conf/comsnets/TerkkiRT17}.

Given our abovementioned observation that ad libraries are not used only for serving ads, further studies of ad libraries need to be careful that the analyzed apps are ad-displaying apps (i.e., the integrated ad libraries are used for serving ads).
Otherwise, researchers on mobile ad libraries could falsely identify the ad-displaying apps.

\begin{table}[!t]
	\centering
	\caption{Statistics of the studied apps.}
	\label{tab:stats_app_serve_ads}
	\resizebox{\columnwidth}{!}{
		\begin{tabular}
		{>{\raggedright\arraybackslash}p{2.4cm}%
        >{\raggedright\arraybackslash}p{5cm}%
        >{\raggedleft\arraybackslash}p{0.8cm}%
        >{\raggedleft\arraybackslash}p{0.8cm}%
        }
			\toprule
			\multicolumn{1}{C{2.3cm}}{\textbf{App category}} &
			\multicolumn{1}{C{5cm}}{\textbf{Category definition}} &
			\multicolumn{1}{C{0.8cm}}{\textbf{\# of apps}} &
			\multicolumn{1}{C{0.8cm}}{\textbf{\% of apps}} \\
			\midrule
			
			Ad-displaying  &Apps that integrate ad libraries and \textbf{display ads} (i.e., apps that call show-ad methods).
    &1,076  &58.6\%\\ \midrule
			Non-ad-displaying  &Apps that \textbf{do not contain} any of the identified ad library packages.        &530    &28.9\% \\ \cmidrule(l){2-4}
			    &Apps that integrate \textbf{Google AdMob for analytical purposes} instead of displaying ads.        &154    &8.4\% \\
			\midrule
			Others &Apps that contain ad library packages that \textbf{are not used (called)} by any other packages in the app: 69 apps with native code and 8 apps that do not contain native code.          &77    &4.2\% \\
			\bottomrule     
		\end{tabular}
	}
\end{table}

\begin{table}[t]
	\centering
	\caption{Top ten third party libraries that depend on the Google AdMob ad library.}
	\label{tab:stats_third_party_libs_use_Ads_Sdk}
	\resizebox{\columnwidth}{!}{
		\begin{tabular}
		{>{\raggedright\arraybackslash}p{4.8cm}%
        >{\raggedright\arraybackslash}p{3.3cm}%
        >{\raggedleft\arraybackslash}p{0.9cm}%
        >{\raggedleft\arraybackslash}p{1.0cm}%
        }
			\toprule
			\multicolumn{1}{C{4.8cm}}{\textbf{Package name}} &
			\multicolumn{1}{C{3.3cm}}{\textbf{Library name}} &
			\multicolumn{1}{C{1.1cm}}{\textbf{\# of apps using the package}} &
			\multicolumn{1}{C{1.2cm}}{\textbf{\% of apps using the package}} \\
			\midrule
			com.google.android.gms.analytics~\cite{google_analytics}	    &Google Analytics	&151	&98.1\%\\
			com.appsflyer~\cite{appsflyer_analytics}	                        & AppsFlyer Analytics	&23	&14.9\%\\
			com.flurry.sdk~\cite{flurry_analytics}	                        & Flurry Analytics	&14	&9.1\%\\
			com.kochava.android.tracker~\cite{kochova_analytics}	            & Kochava Analytics	&13	&8.4\%\\
			com.localytics.android~\cite{localitics_analytics}	                & Android Location Tracker	&10	&6.5\%\\
			com.life360.android.location~\cite{life360_analytics}	        & Life 350 Location Tracker	&4	&2.6\%\\
			com.mologiq.analytics~\cite{ninthdecimal_analytics}	                & MoLogiq Analytic	&4	&2.6\%\\
			com.quantcast.measurement.service~\cite{quantcast_analytics}	    & Quantcast Measure	&4	&2.6\%\\
			com.urbanairship.analytics~\cite{urban_airship_analytics}	            & Urban Airship Analytics	&3	&1.9\%\\
			com.moat.analytics.mobile.ovi~\cite{moat_analytics}	        & Moat Analytics	&2	&1.3\%\\
			\bottomrule     
	\end{tabular}}
\end{table}

\begin{table}[t!]
\centering
\caption{Statistics for the top ten integrated ad libraries.}
\label{tab:top_ad_libs}
\begin{tabular}{lrr}
\toprule
\multicolumn{1}{C{2.5cm}}{\textbf{Ad library}} &
\multicolumn{1}{C{1.5cm}}{\textbf{\# of ad-displaying apps}} &
\multicolumn{1}{C{1.5cm}}{\textbf{\% of ad-displaying apps}} \\
\midrule
Google AdMob	            &1,043	&96.9\%\\
Facebook Audience Network	    &478	&44.4\%\\
MoPub	            &287	&26.7\%\\
Amazon Mobile Ad	&122	&11.3\%\\
Flurry	            &105	&9.7\%\\
InMobi	            &105	&9.7\%\\
Millennialmedia	    &104	&9.6\%\\
AdColony	        &91	    &8.5\%\\
Applovin	        &84 	&7.8\%\\
Unity Ads	        &65	    &6.1\%\\
\bottomrule     
\end{tabular}
\end{table}

\noindent\textbf{Although the \textit{Google AdMob} and \textit{Facebook Audience Network} are the most integrated ad libraries throughout the studied ad-displaying apps, some ad libraries are popular within certain app categories.} 
Table~\ref{tab:top_ad_libs} presents the top ten integrated ad libraries of the studied ad-displaying apps.
The Google AdMob is the most widely integrated ad library (96.4\% of the ad-displaying apps integrate the Google AdMob ad library).

To understand the popularity of an ad library in every app category, we measured the percentage of apps that integrate every ad library in each app category.
Table~\ref{tab:top_five_rank_across_category} shows the top five integrated ad libraries in each app category.
\begin{table}[h!]
\centering
\caption{Top five ranked ad libraries in each app category.}
\label{tab:top_five_rank_across_category}
\resizebox{\columnwidth}{!}{
\begin{tabular}
{>{\raggedright\arraybackslash}p{2.7cm}%
   >{\raggedright\arraybackslash}p{1.1cm}%
   >{\raggedright\arraybackslash}p{1.1cm}%
   >{\raggedright\arraybackslash}p{1.1cm}%
    >{\raggedright\arraybackslash}p{1.1cm}%
     >{\raggedright\arraybackslash}p{1.1cm}%
  }
\toprule
\multicolumn{1}{C{2.7cm}}{\textbf{App category}} &
\multicolumn{1}{C{1.0cm}}{\textbf{Rank 1}} &
\multicolumn{1}{C{1.0cm}}{\textbf{Rank 2}} &
\multicolumn{1}{C{1.0cm}}{\textbf{Rank 3}} &
\multicolumn{1}{C{1.0cm}}{\textbf{Rank 4}} &
\multicolumn{1}{C{1.0cm}}{\textbf{Rank 5}} \\
\midrule
Music and audio	    
&GA(95\%) &FAN(34\%)	&MP(23\%)	&MM(13\%)	&IM(11\%)\\
Weather	            
&GA(100\%)	&\textbf{MP(53\%)}	&FAN(45\%)	&AMA(38\%)	&MM(30\%)\\
Personalization	    
&GA(96\%)	&FAN(85\%)	&MP(37\%)	&FL(9\%)	&UA(8\%)\\
Entertainment	    
&GA(85\%)	& FAN(43\%)	& MP(27\%)	& AV(23\%)	& UA(21\%)\\
Photography	        
&GA(94\%)	&FAN(59\%)	&MP(23\%)	&MV(19\%)	&AV(10\%)\\
Game	            
&GA(90\%)	& \textbf{UA(52\%)}	& AC(50\%)	& VL(40\%)	& AV(41\%)\\
News and magazines	
&GA(95\%)	&FAN(31\%)	&MP(28\%)	&FH(15\%)	&IA(11\%)\\
Tools	            
&GA(98\%)	&FAN(69\%)	&MP(45\%)	&FL(18\%)	&DAP(18\%)\\
Video players	    
&GA(95\%)	&FAN(28\%)	&MP(15\%)	&IM(8\%)	&AV(6\%)\\
Auto and vehicles	
&GA(100\%)	&FAN(14\%)  &MP(14\%)		& \hspace{0.6cm}-- &  \hspace{0.6cm}-- \\
Sports	            
&GA(90\%)	&FAN(20\%)	&FH(15\%)	&MP(13\%)	&MM(11\%)\\
Social	            
&GA(92\%)	&FAN(65\%)	&MP(37\%)	&FL(27\%)	&IM(22\%)\\
Comics	            
&GA(88\%)	&FAN(30\%)	&AMA(22\%)	&AC(19\%)	&IM(13\%)\\
Books and reference	
&GA(88\%)	&FAN(24\%)	&MP(13\%)	&AMA(13\%)	&AB(11\%)\\
Health and fitness	
&GA(100\%)	&FAN(35\%)	&MP(20\%)	&AMA(17\%)	&MV(10\%)\\
Productivity	    
&GA(95\%)	&FAN(64\%)	&MP(26\%)	&FL(16\%)	&DAP(9\%)\\
Lifestyle	        
&GA(100\%)	&FAN(45\%)	&MP(31\%)	&AMA(18\%)	&FL(13\%)\\
Communication	    
&GA(89\%)	&FAN(54\%)	&MP(35\%)	&FL(21\%)	&IM(18\%)\\
Medical	            
&GA(100\%)	&\textbf{MP(30\%)}	&FAN(23\%)	&AM(15\%)	&AC(11\%)\\
Shopping	        
&GA(90\%)	&FAN(18\%)	&TJ(4\%)	&MP(4\%)	&VL(4\%)\\
Finance	            
&GA(100\%)	&FAN(13\%)	&MP(6\%)	&FY(6\%)	&MM(6\%)\\
Maps and navigation	
&GA(92\%)	&FAN(7\%)   &MP(7\%)		&AS(3\%)	& \hspace{0.6cm}-- \\
Travel and local	
&GA(78\%)	&\textbf{MP(21\%)}	&MM(14\%)	&FL(7\%)	&AOL(7\%)\\
Education	        
&GA(96\%)	&FAN(33\%)	&MP(22\%)	&FL(7\%)	& \hspace{0.6cm}-- \\
Libraries and demo	
& GA(65\%)	&\textbf{MP(11\%)}	&IM(11\%)	&FL(11\%)	&MP(11\%)\\
Business	        
&GA(91\%)	&FAN(35\%)	&MP(13\%)	&AMA(8\%)	&DAP(4\%)\\
\bottomrule
\multicolumn{6}{l}{\pbox{30cm}{The abbreviations for ad libraries are as follows: AdColony \textit{(AC)}, AdMarvel \textit{(AM)},}} \\
\multicolumn{6}{l}{\pbox{30cm}{AerServ \textit{(AS)}, Amazon Mobile Ad \textit{(AMA)}, AppBrain \textit{(AB)}, Du Ad Platform \textit{(DAP)},}}\\
\multicolumn{6}{l}{\pbox{30cm}{Facebook Audience Network \textit{(FAN)}, Flurry \textit{(FL)}, FreeWheel \textit{(FH)}, Google AdMob}}\\
\multicolumn{6}{l}{\pbox{30cm}{\textit{(GA)}, InMobi \textit{(IM)}, MillennialMedia \textit{(MM)}, MobVista \textit{(MV)}, MoPub \textit{(MP)}, TapJoy}}\\
\multicolumn{6}{l}{\pbox{30cm}{\textit{(TJ)}, Unity Ads \textit{(UA)}, and Vungle \textit{(VL)}.}}\\
\multicolumn{6}{l}{\pbox{30cm}{* The bold text highlights ad libraries (in Rank 2) other than Facebook Audience}} \\
\multicolumn{6}{l}{\pbox{30cm}{Network \textit{(FAN)} ad library.}} \\
\end{tabular}
}
\end{table}
We observe that the Google AdMob and Facebook Audience Network are the most integrated ad libraries in each app category.  
However, other ad libraries are popular within certain app categories.
For example, we observe that the Unity Ads ad library is the second most integrated ad library in the Game category (52\% of the ad-displaying apps in the Game category integrate the Unity Ads ad library). 
One possible reason for the popularity of the Unity Ads ad library in the Game category is that the library provides easy integration to the apps that are built on the Unity framework (a popular framework for building games).
In addition, the Unity Ads ad library offers features for displaying rewarded video ads (e.g., users earn an extra life or coins if they watch a video ad), which have become popular among video gaming apps as these ads improve user engagement with the app~\cite{Reward_Video_Increase_User_Experience,Reward_Video_Continues_Dominates}.

We also observe that the MoPub (MP) ad library is the second most popular ad library in four app categories (i.e., the Weather, Medical, Travel and local, and Libraries and demo app categories).
One possible reason for the MoPub's popularity in these categories is that the MoPub ad library offers an external-ad-mediator. 
The external-ad-mediator is an ad library feature that facilitates the integration of multiple other ad libraries.
In particular, we observe that 76\% of the ad-displaying apps in the Weather category integrate multiple ad libraries with the external-ad-mediator of the MoPub ad library.
We also observe that the external-ad-mediator of the MoPub ad library is the most used external-ad-mediator in other app categories (i.e., Medical, Travel and local, and Libraries and demo app categories).

\par
\begin{summary}{}{}
	While ad libraries are commonly integrated for serving ads, they are often integrated for analytical purposes.
	The mobile ad market is heavily dominated by the Google AdMob and Facebook Audience Network ad libraries, yet other ad libraries still play a leading role in some particular app categories.
\end{summary}

\section{A Study of the Integration Strategies of Ad Libraries}
\label{sec:Results}
We now present  our study of  the integration strategies of ad libraries. For each research question, we discuss the motivation, approach and results.

\subsection{RQ1: What are the characteristics of apps which integrate multiple ad libraries?}
\label{sec:RQ1_Results}
\noindent\emph{Motivation:}
Ad networks decide whether to serve ads for a requesting app based on different factors (e.g., the characteristics of the user-base of that app).
Hence, apps most often integrate more than one ad library.
A good understanding of the apps that integrate multiple ad libraries would help the developers of ad libraries better understand how apps use their ad libraries and how their libraries co-exist with other competing ad libraries.

\noindent\emph{Approach:}
For this study, we calculated the percentage of ad-displaying apps that integrate a specific number of ad libraries. 
Then, we calculated the \textit{multiple-ads} ratio (as the ratio of apps that integrate  \textit{multiple ad libraries} to apps that integrate \textit{a single ad library}) across every download range. 
A multiple-ads ratio that is higher than one indicates that the number of apps that integrate multiple ad libraries is higher than the number of apps that integrate a single ad library (for a certain download range).

\begin{figure}[!t]
    \centering
    \includegraphics[width=0.5\textwidth]{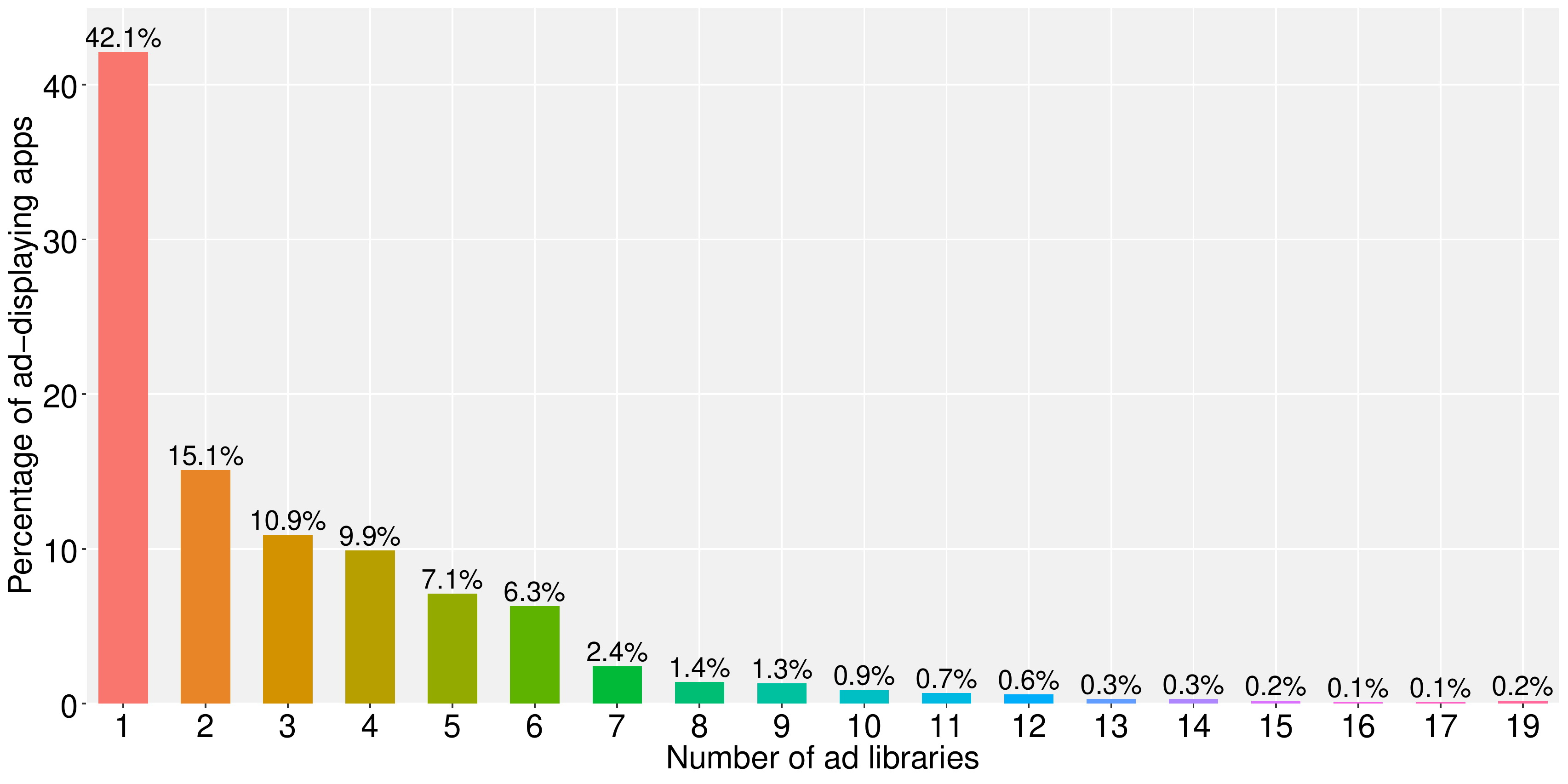}
    \caption{The percentage of ad-displaying apps that integrate a specific number of ad libraries.}
    \label{fig:stats_number_integrated_ad_library}
\end{figure}

\noindent\emph{Findings:}
\textbf{57.9\% of the ad-displaying apps integrate multiple ad libraries.}
Figure~\ref{fig:stats_number_integrated_ad_library} shows the percentage of ad-displaying apps along with the number of integrated ad libraries. 
The number of integrated ad libraries could reach up to 19 ad libraries.
For instance, the ``FreeTone Free Calls \& Texting''\footnote{https://play.google.com/store/apps/details?id=com.textmeinc\\.freetone} app integrates 19 ad libraries; these ad libraries represent 32\% of the binary size of the app.
App developers integrate multiple ad libraries to increase the ad fill rate (i.e., to ensure that their apps can always display an ad)~\cite{israel_app_rating_ads_library}.

\begin{figure}[!t]
    \centering
    \includegraphics[width=0.5\textwidth]{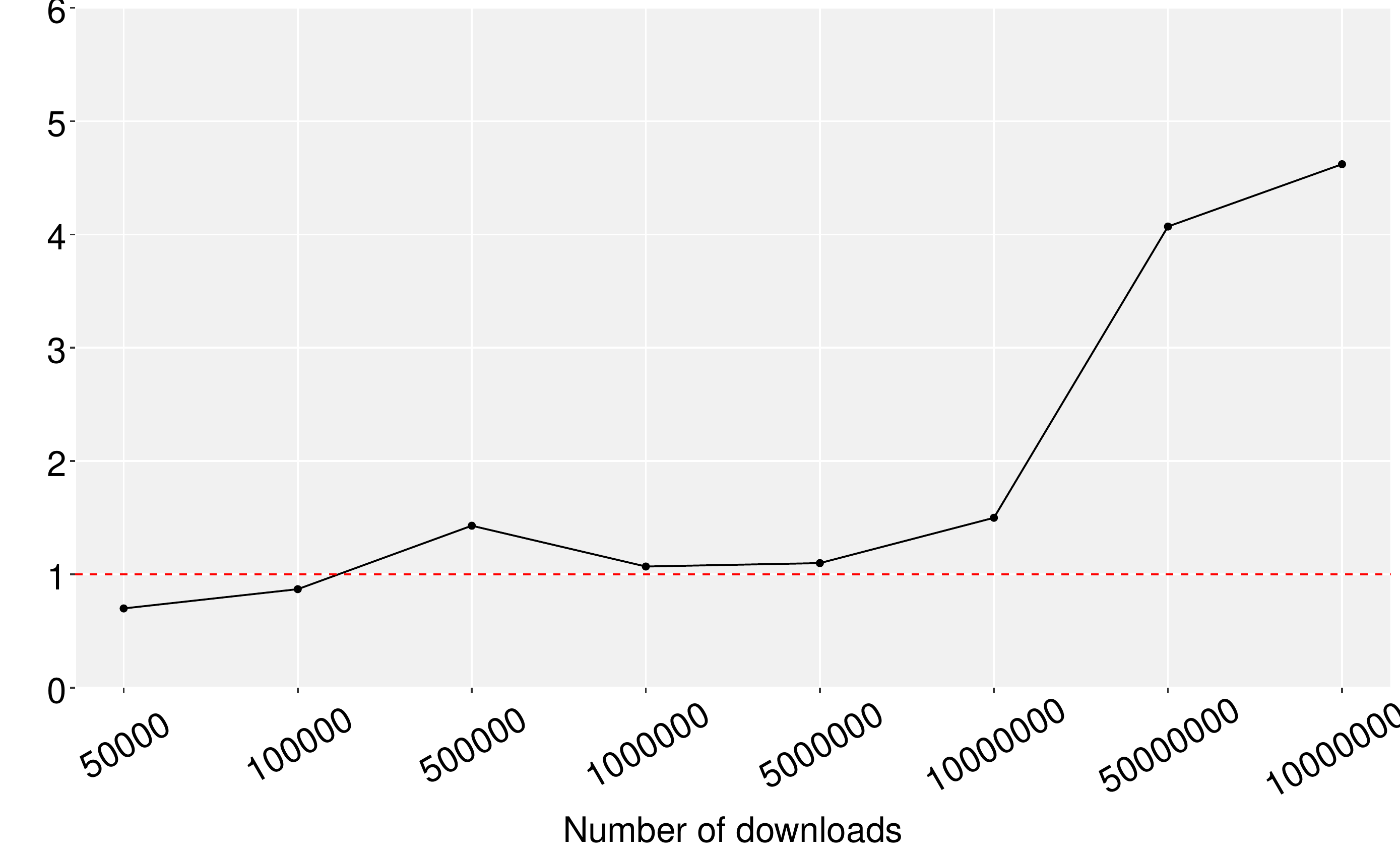}
    \caption{A line plot showing the ratio of the number of apps that integrate more than one ad library over the number of apps that integrate a single ad library  across the number of app downloads. The \textcolor{red}{\textbf{red}} dotted line in the figure shows the ratio value 1.} 
    \label{fig:download_ad_integration}
\end{figure}

\textbf{Apps with a large number of downloads are more likely to integrate multiple ad libraries.}
Figure~\ref{fig:download_ad_integration} presents a line plot of the multiple-ads ratio along with the number of downloads.
The ratio value increases as the number of downloads increases.
The increase in the ratio value from one to five indicates that apps having a high number of downloads tend to integrate multiple libraries.

The Spearman's rank-order correlation between the number of downloads and the multiple-ads ratio is $\rho = 0.92$ (with a $p-value$ less than $0.05$), indicating that the probability of integrating multiple ad libraries increases with the growth in the number of downloads.

\begin{table}[t!]
\centering
\caption{Distribution of apps that use ad libraries in each app category.}
\label{tab:app_use_ad_library_across_app_categories}
\resizebox{\columnwidth}{!}{
\begin{tabular} {>{\raggedright\arraybackslash}p{2.7cm}%
   >{\raggedleft\arraybackslash}p{0.6cm}%
   >{\raggedleft\arraybackslash}p{0.6cm}%
   >{\raggedleft\arraybackslash}p{0.8cm}%
    >{\raggedleft\arraybackslash}p{0.7cm}%
     >{\raggedleft\arraybackslash}p{0.8cm}%
  }

\toprule
\multicolumn{1}{C{2.7cm}}{\textbf{App category}} &
\multicolumn{1}{C{1.0cm}}{\textbf{\# of studied apps}} &
\multicolumn{1}{C{1.0cm}}{\textbf{\# of ad-displaying apps}} &
\multicolumn{1}{C{0.9cm}}{\textbf{\% of ad-displaying apps}} &
\multicolumn{1}{C{1.2cm}}{\textbf{Median \# of integrated ad libraries}}& 
\multicolumn{1}{C{1.3cm}}{\textbf{Maximum \# of integrated ad libraries}}\\ 
\midrule
Music and audio	        &67	    &64	    &95\% &2 &11\\
Weather	                &76	    &71	    &93\% &5 &13\\
Personalization	        &89	    &82	    &92\% &3 &10\\
Entertainment	        &52	    &42	    &81\% &3 &15\\
Photography	            &94	    &77	    &81\% &2 &10\\
Game	                &59	    &47	    &79\% &6 &17\\
News and magazines	    &78	    &59	    &76\% &2 &7\\
Tools	                &93	    &69	    &74\% &3 &9\\
Video players	        &59	    &44	    &71\% &1 &7\\
Auto and vehicles	    &10	    &7	    &70\% &1 &2\\
Sports	                &74	    &52	    &70\% &2 &11\\
Social	                &81	    &54     &67\% &3 &19\\
Comics	                &56	    &36	    &64\% &2 &7\\
Books and reference	    &77	    &45	    &58\% &1 &9\\
Health and fitness	    &70	    &38	    &54\% &2 &6\\
Productivity	        &80	    &42	    &52\% &2 &10\\
Lifestyle	            &42	    &20	    &49\% &2 &7\\
Communication	        &76	    &36	    &48\% &3 &12\\
Medical	                &55	    &26	    &47\% &1 &6\\
Shopping	            &53	    &22	    &42\% &1 &4\\
Finance	                &38	    &15	    &39\% &1 &4\\
Travel and local	    &69	    &27	    &39\% &1 &8\\
Maps and navigation	    &69	    &26 	&37\% &1 &3\\
Education	            &70	    &24	    &34\% &1 &6\\
Libraries and demo	    &27	    &9	    &33\% &1 &5\\
Business	            &81	    &23 	&28\% &1 &5\\
\bottomrule     
\end{tabular}
}
\end{table}
\textbf{Ad-displaying apps are distributed across app categories -- with apps in categories having a high proportion of ad-displaying apps integrating multiple ad libraries.}
Table~\ref{tab:app_use_ad_library_across_app_categories} presents the distribution of ad-displaying apps, the median number of integrated ad libraries, and the maximum number of integrated ad libraries in each app category.
Ad-displaying apps are distributed across app categories, with some categories having a higher penetration of ads.
For example, more than 90\% of the studied apps in the \textit{Music and audio}, the \textit{Weather}, and the \textit{Personalization} app categories integrate ad libraries.

The median number of integrated ad libraries is greater than one for 57.6\% of the app categories.
The Spearman's rank-order correlation between the percentage of ad-displaying apps and the number of integrated ad libraries (median) for every app category is $\rho = 0.7$ ($p-value$ is less than $0.05$), indicating that the number of integrated ad libraries increases with the growth in the proportion of ad-displaying apps within a category.

One of the possible explanations for integrating multiple ad libraries is that as the number of downloads of an app increases or the proportion of ad-displaying apps increases in a category, the competition for ads to display from ad libraries increases, which in turn leads to a lower fill rate.
Hence, integrating multiple ad libraries increases the chance of having an ad to display and the potential ad revenue for ad-displaying apps~\cite{israel_app_rating_ads_library}.

\begin{Summary}{}{}
The probability of integrating multiple ad libraries increases as the number of downloads of an app increases. 
Apps in categories with a high proportion of ad-displaying apps are more likely to integrate multiple ad libraries.
We hypothesize that the integration of multiple ad libraries is a mechanism to cope with the high demand for ads in an attempt to improve the ad fill rate.
\end{Summary}

\subsection{RQ2: How do app developers integrate multiple ad libraries?}
\label{sec:RQ2_Results}
\noindent\textit{Motivation:}
Integrating multiple ad libraries is a common practice in ad-displaying apps.
A good understanding of how app developers integrate multiple ad libraries can help ad library developers identify the challenges and the possible improvements for their ad libraries.

\noindent\textit{Approach:} 
To identify the strategies for integrating multiple ad libraries, the first and the second author of this paper manually analyze several ad-displaying apps where app developers integrate multiple ad libraries as follows.

\noindent \textbf{Step 1: Selecting a statistically representative sample of ad-displaying apps.}
Our data set has 623 ad-displaying apps that integrate multiple ad libraries.
Analyzing all these ad-displaying apps manually is both difficult and time consuming.
Therefore, for our manual study, we selected a statistically representative random sample of 62 apps (out of the 623 ad-displaying apps) providing us with a confidence level of 90\% and a confidence interval of 10\%.

\noindent \textbf{Step 2: Generating a static call graph for each selected app.}
To understand how app developers integrate multiple ad libraries, we need to analyze the \textit{\textbf{call-site}} source code (i.e., the packages, classes, and methods that are needed to communicate with an ad library).
Hence, we decompiled the generated JARs (in Section~\ref{fig:methodology}) into Java source code files using the Class File Reader (CFR) tool~\cite{CFR_TOOL}.
Then, we used the Understand tool~\cite{understand_tool} to generate and visualize the dependency call graph of each studied ad-displaying app.

\noindent \textbf{Step 3: Identifying the strategies of integrating multiple ad libraries.}
We start our manual analysis with an open-ended question ``How does an app integrate multiple ad libraries?''.
We observe that apps differ in the way by which they integrate ad libraries with respect to two practices:
(1) whether the app code uses a centralized component (i.e., an ad \textbf{mediator component}) that handles the access to the multiple ad libraries 
and (2) whether the centralized component is written by the app developer or by the library designer.

Hence, we manually investigate every selected ad-displaying app based on the following two questions:
``Does the app code call a centralized component that handles the access to the multiple ad libraries?''
and ``Is that centralized component written by the app developer or by the library developer?''.
Then, we grouped apps with similar integration behavior (in the context of the aforementioned investigated questions) as an \textbf{integration strategy}.
Finally, we derived a set of rules to automatically identify the integration strategy for any unseen app.

To generate the static call graph and analyze the app code (i.e., classes, methods, and packages) for identifying integration strategies, we took 40 minutes on average for each of the studied apps. 

\noindent \textbf{Step 4: Analyzing the characteristics of the identified integration strategies.}
We ran the derived rules on the studied 623 ad-displaying apps and identified apps that belong to every integration strategy.
Then, we studied the characteristics (i.e., the number of call-cite classes) of the apps that belong to every integration strategy.
Finally, based on our analysis of the apps, we provide a description, an example, the benefits, and the drawbacks of each identified strategy for integrating multiple ad libraries.

To better understand the integration strategies for ad libraries, we analyze qualitative data sources. In particular, we manually examine two artifacts: (1) 500 Stack Overflow (SO) posts that are related to mobile ads, and (2) 35 ad-related articles from the developer forums and blogs of the top ten ad libraries (e.g., the InMobi blog) as follows:

\textbf{Step 1: Collecting qualitative data.} To identify the Android ad-network-related posts in Stack Overflow we employ SO's search option using the following keywords: ``multiple ad network'', ``fill rate android'', ``ad mediation'', and ``admob banner facebook native''. The SO's search option returned 255 posts for ``multiple ad networks'', 118 posts for ``file rate android'', 113 posts for ``mediation ads android'', and 14 posts for ``admob banner facebook native''. In total, we collect 500 ad-network-related posts. The collected SO posts are answered within one (median) day with a median of 667 views  -- highlighting that our topic and addressed challenges are of great relevance to the development community.
To identify ad-network-related articles, we search on Google using a combination of the aforementioned keywords and the names of the different ad libraries. We select only those articles that are related to the forum discussions (e.g., Google Mobile Ads SDK Technical Forum) and blogs of ad libraries. We collected 35 ad-related articles from the developer forums and blogs of the top ten ad libraries.

\textbf{Step 2: Investigating qualitative data.} In this step, the first two authors of this paper manually investigate each of the collected SO posts. Our objective of the analysis is to understand the strategies that developers follow to integrate multiple ad libraries into their apps and the associated issues with these strategies. To achieve this objective, we carefully read the title and body of each SO post. In addition, we examine the answers and the comments of every post to understand the integration strategies for ad libraries. Although each of the selected SO posts adds knowledge about integrating multiple ad libraries, we identified 25 SO posts that specifically mention drawbacks or benefits of the integration strategies for ad libraries.

During our manual analysis, if there was a disagreement in the meaning of a post, we (the authors) carefully reread the answers and the comments related to the post and further discussed it until consensus was reached. Since both authors analyzed together all SO posts and agreements (on all analysis of the examined posts) were reached in the end, the authors did not compute the inter-rater agreement. To facilitate the replicability of our work, the studied SO posts are available in our replication package. We follow the same approach to manually study the 35 ad-related articles to understand the process of integrating multiple ad libraries.

It took around 7 minutes to read each SO post and 10 minutes to read each ad-related article (from each author). In total, it took around 12 hours from each author to finish analyzing all the selected SO posts and ad-related articles.

\vspace{0.1cm}
\noindent\emph{Findings:}
We identified four strategies for integrating multiple ad libraries. 
Table~\ref{tab:stats_integration_strategy_multiple_ad_library} shows the distribution of apps across the identified strategies.
In the next section, we explain the identified integration strategies.

\begin{table}[t!]
	\centering
	\caption{Distribution of apps across the different integration strategies.}
	\label{tab:stats_integration_strategy_multiple_ad_library}
	\begin{tabular}{lrr}
		\toprule
		\multicolumn{1}{C{3cm}}{\textbf{Ad library integration strategy}} &
		\multicolumn{1}{C{1.7cm}}{\textbf{\# of apps}} &
		\multicolumn{1}{C{1.7cm}}{\textbf{\% of apps}}\\ \midrule
		Mixed strategy	            &317	    &50.9\%\\
		Self-mediation strategy	    &166	    &26.6\%\\
		External-mediation strategy	&63	        &10.2\%\\
		Scattered strategy	    &77	    &12.3\%\\
		\bottomrule     
	\end{tabular}
\end{table}

\begin{figure}[!t]
	\centering
	\includegraphics[width=0.4\textwidth]{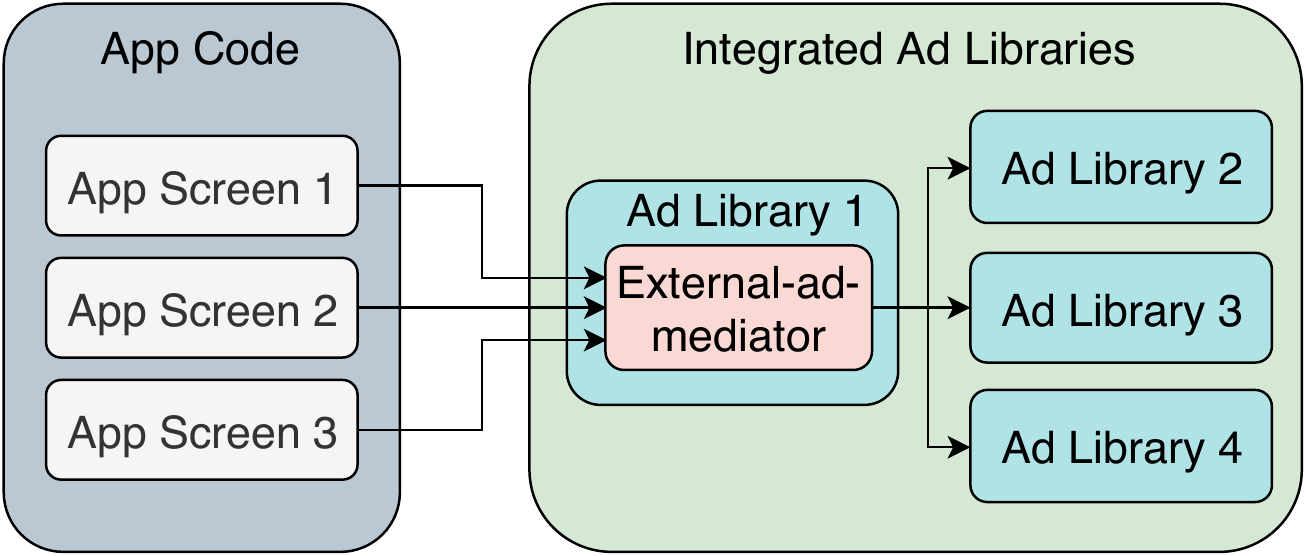}
	\caption{An overview of the external-mediation strategy.} 
	\label{fig:rq2_external_mediation_strategy}
\end{figure}


\bigskip
\noindent \textbf{{\Large (1) External-mediation strategy}}
\bigskip

\noindent \textbf{Description of the external-mediation strategy:}

\noindent In this strategy, app developers write code to integrate only one ad library that offers a mediator package. 
This mediator package is responsible for serving ads from other ad networks, which are supported by the ad library.
Since the mediator is not written by the app developers, we call this an external-mediation strategy.

Figure~\ref{fig:rq2_external_mediation_strategy} shows an overview of how app developers integrate multiple ad libraries using an external-mediation strategy. 
App developers integrate an ad library (Ad Library 1) that has an external-ad-mediator. 
Every app screen that displays ads communicates only with the external-ad-mediator of the Ad Library 1. 
The external-ad-mediator communicates with the integrated ad libraries and serves ads from these libraries.

\noindent \textbf{Rules for automatically identifying apps that use the external-mediation strategy:}

\noindent We determine that an ad-displaying app is using the external-mediation strategy if the following two rules are met:

\begin{enumerate}[wide, labelwidth=!, labelindent=0pt]
    \item  The number of accessed ad libraries by the app code is one, and the number of integrated ad libraries in the app is more than one.
    \item The package of the accessed ad library contains an external-ad-mediator package that is accessed by the app code.
\end{enumerate}


\noindent \textbf{An example app that uses the external-mediation strategy:}

\noindent The \textit{``Ringtones \& Wallpapers for Me''}\footnote{https://play.google.com/store/apps/details?id=com.apalon.\\ringtones} app, a popular app in the Personalization category, displays ads from ten ad libraries. 
The app code (i.e., the code that is written by the app developer) of this app contains the code for integrating only one ad library (Google AdMob).
This app uses the external-ad-mediator of the Google AdMob ad library (com.google.andorid.gms.ads.mediation) which communicates with the other nine ad libraries as well as the Google AdMob ad library for displaying ads.

\noindent \textbf{Benefits of using the external-mediation strategy:}

\begin{itemize}[wide, labelwidth=!, labelindent=0pt]
       \item The ease of integrating multiple ad libraries is one of the main benefits of this strategy as the external-ad-mediator implements the required logic for serving ads from multiple ad libraries~\cite{tapjoy_mobile_ad_mediation}. For example, to integrate a new ad library, app developers include the new ad library (along with the external-ad-mediator) in their apps. This process does not require any changes to the ad-call-cite code (compared to the self-mediator strategy that needs changes to the self-mediation support code). The ease of integrating multiple ad libraries maybe one of the reasons for the high ratio of adding or deleting ad libraries for the external-ad-mediator strategy compared to the other integration strategies.
	
        Nevertheless, based on our qualitative analysis of Stack Overflow questions, we noticed that the process of using the external-mediation strategy is not that intuitive as developers may not know that they need to include the external-ad-mediator in their apps (in addition to adding the required ad libraries)\footnote{ https://stackoverflow.com/questions/14481380/how-implement-the-mediation-ad-in-android}\footnote{ https://stackoverflow.com/questions/48363760/admob-mediation-with-facebook-audience-network-in-xamarin-forms}\footnote{ https://stackoverflow.com/questions/25008446/android-mediation}.
        
        \item The external-ad-mediator selects an ad library for serving ads from the supported ad libraries based on dynamically estimated measures such as the eCPM which captures the ad monetization performance of an ad library at run-time; leading to much more dynamic and accurate estimates of the revenue for a served ad~\cite{definition_ecpm}.
    	
\end{itemize}


\noindent \textbf{Drawbacks of using the external-mediation strategy:}
\begin{itemize}[wide, labelwidth=!, labelindent=0pt]

	\item The external-ad-mediator of an ad library may not support all the existing ad libraries that are available in the app market, and integrating unsupported ad libraries could crash mobile apps~\cite{Gabriel_api_change_fault_proneness}. 
    Therefore, app developers cannot serve ads from the unsupported ad libraries.
	
	For example, we observe that only 5 out of the identified 63 ad libraries offer an external-ad-mediator. 
	The external-ad-mediators of these five ad libraries (Google AdMob, MoPub, AerServ, Fyber and HeyZap) offer support for serving ads from only 13 ad libraries (20\% of the identified ad libraries).
	Hence, apps that use the external-mediation strategy cannot serve ads from other ad libraries unless they are supported by the external-ad-mediators.
	
	\item  The entire process of serving an ad is not transparent as app developers have less control over the exact ad library from which an ad is to be served. 
    For example, in one of the discussions in the Google Mobile Ads SDK Technical Forum about how external-ad-mediator works, a developer of the Google Mobile Ads SDK Team states: ``Be noted that AdMob will be the one that would decide which mediated ad would display on an ad unit at any given time, depending on various factors (mostly on network priority and eCPM floors)''~\cite{google_admob_drawback_external_mediation}. This answer indicates that app developers have no control over the selection algorithm.
    In addition, the external-ad-mediator might provide a preferential serving of ads from its network over other ad networks. This lack of transparency might cause a mistrust issue leading app developers to avoid the use of the external-ad-mediators of some ad libraries~\cite{quora_mediation_non_mediation}.

\end{itemize}

\bigskip




\noindent \textbf{{\Large (2) Self-mediation strategy}}
\bigskip

\noindent \textbf{Description of the self-mediation strategy:}

\noindent In this strategy, app developers write their centralized package (i.e., self-mediator), which communicates with the integrated ad libraries and manages the process of serving ads. 

\begin{figure}[!t]
	\centering
	\includegraphics[width=0.4\textwidth]{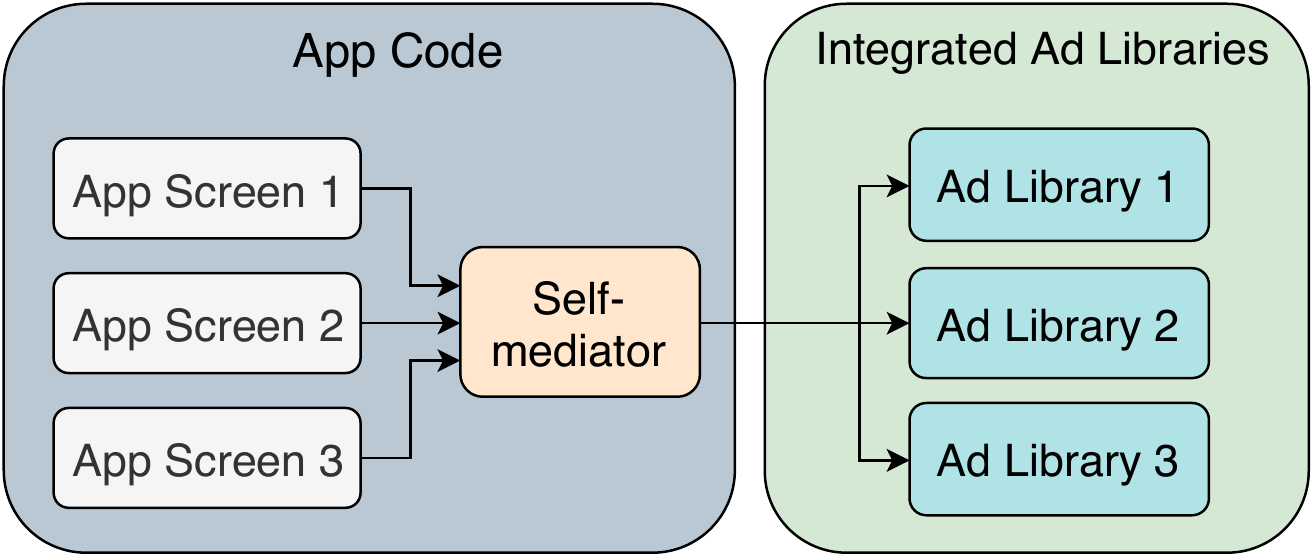}
	\caption{An overview of the self-mediation strategy.} 
	\label{fig:rq2_self_mediation_strategy}
\end{figure}

Figure~\ref{fig:rq2_self_mediation_strategy} presents an overview of the self-mediation strategy. All three app screens communicate with the self-mediator, and the self-mediator communicates with the integrated ad libraries to serve ads from them.

\noindent\textbf{Rules for automatically identifying apps that use the self-mediation strategy:}

\noindent We determine that an ad-displaying app is using the self-mediation strategy if the following rules are met:
\begin{enumerate}[wide, labelwidth=!, labelindent=0pt]
	\item  The number of integrated ad libraries is more than one.
	\item The number of accessed ad libraries by the app code and the number of integrated ad libraries are equal.
	\item The app code contains a centralized package that communicates with the integrated ad libraries.
\end{enumerate}


\noindent \textbf{An example app that uses the self-mediation strategy:}

\noindent The ``\textit{Calculator Plus Free}''\footnote{https://play.google.com/store/apps/details?id=com.digitalchemy.\\calculator.freedecimal} app, a popular app in the Tool category, integrates seven ad libraries using the self mediation strategy. The developer of the app wrote their self-mediator (com.digitalchemy.foundation.advertising), which communicates with the integrated ad libraries to serve ads.


\noindent \textbf{Benefits of using the self-mediation strategy:}
\begin{itemize}[wide, labelwidth=!, labelindent=0pt]
    \item A self mediation strategy provides a good encapsulation of the code code~\cite{book_design_pattern} because app developers write a centralized package that manages the selection and serving of ads from multiple ad libraries.
    
    \item App developers are free to integrate any ad library instead of being limited to a handful of supported ad libraries like in the case of the external-mediation strategy. For example, we observe that apps that use the self-mediation strategy integrate 41 ad libraries with 67.6\% of these libraries not supporting external ad-mediators.

    
    \item App developers have more control over selecting the ad library from which to serve an ad. 
    We observe that app developers mainly use the following three approaches:
    \begin{enumerate}[wide, labelwidth=!, labelindent=0pt]
    
    \item \textbf{A round-robin approach without a preferred list of ad libraries.} In this approach, a random list of the integrated ad libraries is generated. 
    If the first ad library in the list fails to serve an ad, the self-mediator requests an ad from the next ad library. 
    This process continues in a circular order until an ad is served from an ad library.
    
    \item \textbf{A round-robin approach with a preferred list of ad libraries.} In this approach, app developers set a preferred list of the integrated ad libraries based on some measures (e.g., the popularity of the ad library in a country or the offered feature of the ad library). 
    The self-mediator selects the best preferred ad library for requesting an ad. If that library cannot serve an ad, then the mediator selects another ad library in a circular fashion from the preferred list until an ad is served from an ad library.
    
    \item \textbf{A custom event-based approach.} In this approach, the self-mediator of an app selects an ad library based on custom events (e.g., when a user clicks a particular menu item or when a user earns a reward in the app). 
    This custom event-based approach allows app developers to select the most suitable ad library for increasing user-engagement for that particular event.
    
    \end{enumerate}
\end{itemize}


\noindent \textbf{Drawbacks of using the self-mediation strategy:}
\begin{itemize}[wide, labelwidth=!, labelindent=0pt]
    \item App developers must write and maintain the code for the self-mediator. The self-mediator represents 8\% (median) of the total number of classes of an app (in our studied apps). 
    For example, the ``\textit{High-Powered Flashlight}''\footnote{https://play.google.com/store/apps/details?id=com.ihandysoft.\\ledflashlight.mini} app, a popular app in the Tool category, serves ads off 10 ad libraries which are integrated using the self-mediation strategy. 
    The self-mediator (the ``com.ihandysoft.ad'' package) represents 22\% of all the classes of this app. We observe that app developers modify the self-mediation support code in 64.5\% (median) of their updates -- supporting our intuition about the maintenance challenges of the self-mediation strategy.
    
     \item The ordering of ad libraries is static in nature. In contrast, the ordering of the external-ad-mediator is much more dynamic as it can order ad libraries based on the dynamically estimated eCPM value which is calculated at run-time based on the buying and selling of ads as conducted through real-time auctions that are facilitated by digital marketplaces (i.e., ad exchanges)~\cite{mediation_inmobi}. For example, in a  Stack Overflow question about the best practice for implementing a self-mediator for displaying ads\footnote{https://stackoverflow.com/questions/26685425/best-coding-practice-for-implementing-not-using-mediation-multiple-ad-networks}, the accepted answer recommends iterating among the integrated ad libraries (i.e., selecting the first ad library then the next one without considering the current ad network data such as ad fill rate, ad response time, and eCPM). Hence, app developers may not select the best ad library based on the current market conditions~\cite{mediation_inmobi}.

\end{itemize}
\bigskip
\noindent \textbf{{\Large (3) Scattered strategy}}
\bigskip

\noindent \textbf{Description of the scattered strategy:}

\noindent In this strategy, app developers neither write their own mediator nor use the external-ad-mediator of an ad library to serve ads from the integrated ad libraries. Rather, developers write code individually for each app screen to integrate each ad library for that particular screen.

\begin{figure}[!t]
	\centering
	\includegraphics[width=0.35\textwidth]{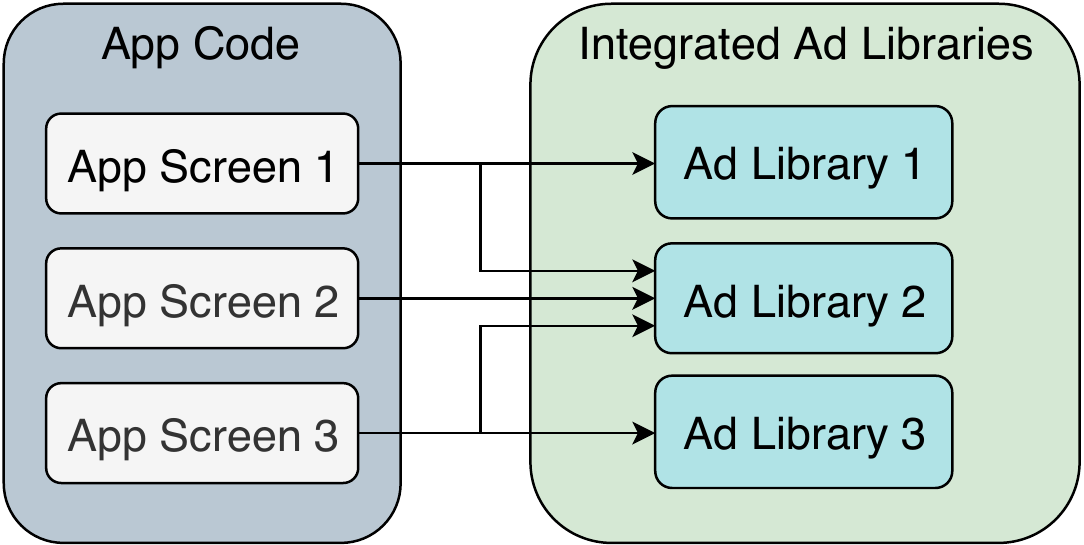}
	\caption{An overview of the scattered strategy.} 
	\label{fig:rq2_scatter_strategy}
\end{figure}

Figure~\ref{fig:rq2_scatter_strategy} shows an overview of the scattered strategy. 
Each app screen communicates with the integrated ad libraries directly. 
The integration code for the Ad Library 2 is written by the app developer for every app screen that displays an ad.

\noindent\textbf{Rules for automatically identifying apps that use the scattered strategy:}

\noindent We determine that an ad-displaying app is using the scattered strategy if the following rules are met:
\begin{enumerate}[wide, labelwidth=!, labelindent=0pt]
    \item The number of integrated ad libraries is more than one.
	\item  The number of accessed ad library by the app code and the number of integrated ad libraries are equal.
	\item The app code does not contain any centralized package that communicates with the integrated ad libraries.
\end{enumerate}

\noindent \textbf{An example app that uses the scattered strategy:}

\noindent The ``\textit{Audiomack -- Download New Music}''\footnote{https://play.google.com/store/apps/details?id=com.audiomack} app, a popular app in Music \& audio category, integrates four ad libraries using the scattered strategy. 
The app displays ads on two screens. 
The developer of this app wrote code in two app screens for displaying ads from ad libraries individually.


\noindent \textbf{Benefits of using the scattered strategy:}
\begin{itemize}[wide, labelwidth=!, labelindent=0pt]
    \item App developers can quickly integrate several ad libraries as developers do not need to write a centralized package (e.g., the self-mediation support code). Designing a flexible and reusable self-mediator is challenging. For example, in a Stack Overflow question about the design of a self-mediator using the Factory pattern\footnote{https://stackoverflow.com/questions/35146989/admodule-architecture-using-abstract-factory-pattern}. The answerer notes the complexity of developing a self-mediator: \textit{``In case some of ads controller need additional action (for example update its state or something) you have to add a new method to interface, and this will be unused with other 100500 implementations''}.
    
     \item App developers can select ads of different ad formats (e.g., banner or native ad format) from different ad libraries based on custom events in their apps. For example, the ``\textit{The Coupons App}''\footnote{https://play.google.com/store/apps/details?id=thecouponsapp.\\coupon} app, a popular app in the Shopping category, integrates Google AdMob and Facebook Audience Network ad libraries in the same app screen. The app selects the Google ad library to serve banner ads and the Facebook Audience Network to serve native ads based on custom events in the app (e.g., the clicking of different buttons in that particular screen).
\end{itemize}


\noindent \textbf{Drawbacks of using the scattered strategy:}
\begin{itemize}[wide, labelwidth=!, labelindent=0pt]

    \item App developers need more effort to maintain their code because they need to write the same integration code (i.e., copy and paste) for an ad library if the ad library is integrated for displaying ads in different app screens. We observe that the probability of modifying an ad-call-site code (i.e., the app code that invokes the methods for integrating an ad library) is 20\% (median) across all ad integration strategies, with that probability increasing considerably to 30\% (an increase of 50\%) for the scattered strategy. We also observe that the median probability of modifying the ad-call-site code for apps that use the mixed strategy is twice that of the median probability of modifying ad-call-site code for apps that use the external-mediation strategy. Hence, app developers need to modify the ad-call-site code at a much larger rate. We further discuss the maintenance effort of each integration strategy in Section~\ref{sec:Discussions}.
    
    \item The scattered code fetches ads from a single ad library for displaying ads on an app screen. For example, the ``\textit{The Coupons App}'' app, a  popular app in the  Shopping category, uses Google AdMob to serve banner ads and uses the Facebook Audience Network to serve native ads in the same app screen. If any of the integrated ad library fails to fetch an ad (e.g., due to the network not filling the ad request), the app screen will fail to display some ads. Alternatively, the app could have used an external-ad-mediator, which provides the needed logic to display ads from different ad libraries if an ad library fails to fill an ad request~\cite{tapjoy_mobile_ad_mediation}. Hence, the scattered strategy is not able to deal with low fill rate issues that might arise.
    
\end{itemize}
\bigskip
\noindent \textbf{{\Large (4) Mixed strategy}}
\bigskip

\noindent \textbf{Description of the mixed strategy:}

\noindent In this strategy, app developers combine both the external-mediation strategy and the scattered strategy to serve ads.

\begin{figure}[!t]
	\centering
	\includegraphics[width=0.4\textwidth]{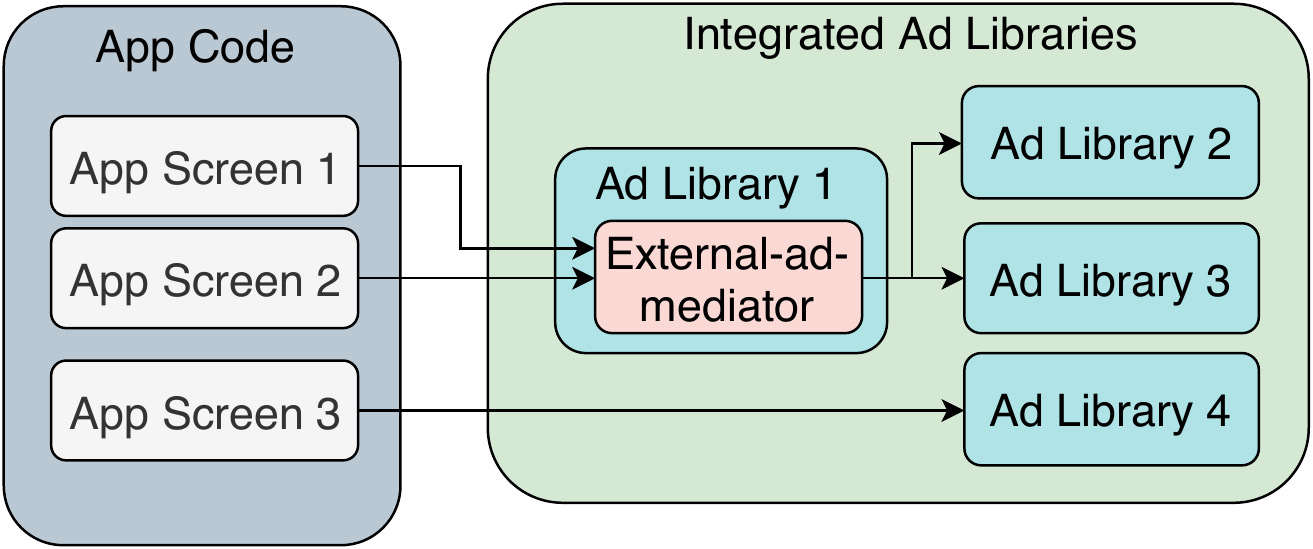}
	\caption{An overview of the mixed strategy.} 
	\label{fig:rq2_mixed_strategy}
\end{figure}
Figure~\ref{fig:rq2_mixed_strategy} shows an overview of the mixed strategy. App Screen 1 and App Screen 2 communicate with ad libraries using an external-mediation strategy, whereas App Screen 3 communicates with the Ad Library 4 using a scattered strategy.

\noindent\textbf{Rules for automatically identifying apps that use the mixed strategy:}

\noindent We determine that an ad-displaying app is using a scattered strategy if the following rules are met:
\begin{enumerate}[wide, labelwidth=!, labelindent=0pt]
    \item The number of integrated ad libraries is more than one.
    \item  The number of accessed ad libraries by the app is less than the number of integrated ad libraries.
    \item The app contains an external-ad-mediator package that communicates to many of the integrated ad libraries.
\end{enumerate}


\noindent \textbf{An example app that uses the mixed strategy:}

\noindent The ``\textit{Real Guitar Free - Chords, Tabs \& Simulator Games}''\footnote{https://play.google.com/store/apps/details?id=com.gismart.guitar} app, a popular app in the Music category, integrates 14 ad libraries to serve ads. The developers of the app use the external-ad-mediator of the Google AdMob (com.google.~\allowbreak~android.gms.ads.mediation) ad library to integrate 11 ad libraries.
To integrate the remaining three ad libraries, the app developers write code in the activities of some specific screens using the scattered strategy.

\begin{table}[!t]
	\centering
	\caption{Mean and five-number summary of each strategy for integrating multiple ad libraries.}
	\label{tab:stats_ad_integration_strategy}
	\centering
	\resizebox{\columnwidth}{!}{
		\begin{tabular}{lrrrrrr}
			\toprule
			\multicolumn{1}{C{2.6cm}}{\textbf{Ad library \newline integration strategy}} &
			\multicolumn{1}{C{0.8cm}}{\textbf{Mean}}                            & 
			\multicolumn{1}{C{0.7cm}}{\textbf{Min.}}                            &
			\multicolumn{1}{C{0.5cm}}{\textbf{1st Qu.}}                         &
			\multicolumn{1}{C{0.8cm}}{\textbf{Median}}                          &
			\multicolumn{1}{C{0.5cm}}{\textbf{3rd Qu.}}                         &
			\multicolumn{1}{C{0.7cm}}{\textbf{Max.}}             \\\midrule
			External-mediation strategy                      & 4   & 2 & 2  & 4  & 5       & 10     \\
			Mixed strategy     & 5   & 2 & 4  & 5  & 6     & 19     \\
			Self-mediation strategy                  & 3   & 2 & 2  & 3  & 4       & 12     \\
			Scattered strategy  & 2   & 2 & 2  & 2  & 4     & 5     \\
			\bottomrule
	\end{tabular}}
\end{table}
\noindent \textbf{Benefits of using the mixed strategy:}

\begin{itemize}[wide, labelwidth=!, labelindent=0pt]

    \item In the mixed strategy, app developers can leverage the external-ad-mediator of an ad library to serve ads from different ad libraries and can also write their integration code for other ad libraries that are not supported by the external-ad-mediator. We observe that 73.9\% of ad-displaying apps with the mixed integration strategy call at least one ad library that is not supported by the currently available external-ad-mediators.
    
    \item Apps that use the mixed strategy integrate more ad libraries than the apps that use other strategies. Table~\ref{tab:stats_ad_integration_strategy} shows the mean and five-number summary of the integrated ad libraries for each of the four identified strategies. App developers integrate a maximum of 19 ad libraries using the mixed integration strategy. We find two apps from the \textit{``TextMe, Inc''} company that integrate 19 ad libraries. We observe that they use the external-ad-mediator of the Google AdMob ad library which supports 13 ad libraries, the rest of the ad libraries are integrated using a scattered strategy since they are not supported by the external-ad-mediator of the Google AdMob ad library.
    
    \item One possible reason for using the mixed strategy is that app developers can display ad formats (e.g., native ads) that are not supported by the external-ad-mediator. For example, a Stack Overflow post notes that a developer used an external-ad-mediator to successfully display banner and interstitial ads from both FAN and Google AdMob ad libraries. Later, the developer wanted to display native ads from the Facebook Audience Network (FAN) using the external-ad-mediator of the Google AdMob. The accepted answer indicates that displaying native ads is currently not supported by the external-ad-mediator (\textit{``Mediation through FAN for Native Express Ads is currently not possible. Only Banner ads and Interstitials have been enabled for mediation for FAN.''})\footnote{https://stackoverflow.com/questions/37648710/facebook-audience-network-native-ads-via-admob-mediation-adapter}. Therefore, app developers need to use the mixed strategy to display banner, native, and interstitial ads from multiple ad libraries.
\end{itemize}

\noindent \textbf{Drawbacks of using the mixed strategy:}
\noindent Since the mixed strategy is the combination of the external-mediation strategy and the scattered strategy, some of the drawbacks of these two strategies exist in the mixed strategy. For example, developers need to spend considerable effort on maintaining their ad library code as we observe that the mixed strategy has the highest probability of modifying ad-call-site code (37\%) compared to other integration strategies. We further discuss the maintenance effort of each integration strategy in Section~\ref{sec:Discussions}.

\begin{Summary}{}{}
App developers dominantly use the mixed and the self-mediation strategy to integrate multiple ad libraries. This might be due to the currently available external-ad-mediators not satisfying their needs.
To have more control over selecting ad libraries for displaying ads, app developers write their own centralized packages (self-mediator) based on preferred metrics (e.g., location information) or custom app events in the self-mediation strategy.
\end{Summary}

\section{Discussion of the maintenance overhead of the integrated ad libraries  for each integration strategy}
\label{sec:Discussions}
In this section, we discuss how app developers maintain their integrated ad libraries over time across the different ad library integration strategies. 
In particular, we discuss the modifiability of the ad-call-site code (i.e., the app code that invokes the methods for integrating an ad library) and the flexibility of integrating ad libraries for each integration strategy.

\subsection{The modifiability of ad-call-site code}
In this section, we discuss the modifiability of the ad-call-site code along two aspects: (1) how frequently (in terms of the proportion of the updates of an app) do app developers modify the ad-call-site code, and (2) what is the proportion of the ad-call-site code that is modified across the integration strategies.

To determine if an ad-call-site code is modified, we follow the same approach of Ruiz et al.~\cite{updates_ad_library}. 
In this approach, for each update of an ad-displaying app that integrates multiple ad libraries, we generate the class signatures for all ad-call-site code (we consider only the statements that invoke ad library methods) of the integrated ad libraries.
The ad-call-site code is modified in the app update ($U_{i+1}$) if the signature of the app update ($U_{i+1}$) is different than the signature of the app update ($U_{i}$). 

\textbf{The probability of modifying the ad-call-site code is 20\% (median) across all integration strategies, with that probability increasing
considerably to 37\% (an increase of 60\%) for ad-displaying apps which use the mixed strategy.}
\begin{figure}[!t]
	\centering
	\includegraphics[width=0.45\textwidth]{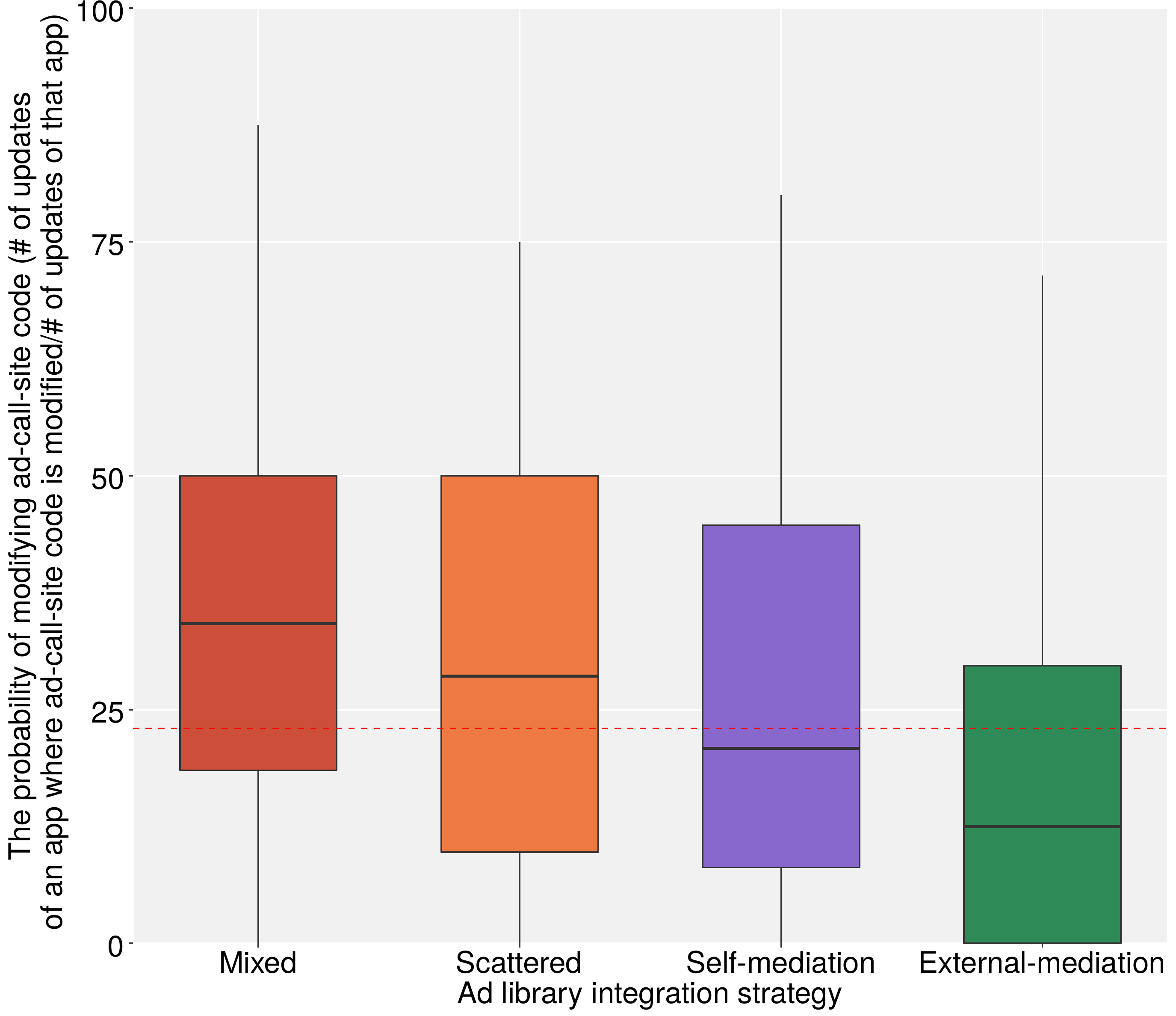}
	\caption{The probability of modifying the ad-call-site code in a update for each ad library integration strategy. The {\color{red}red}~dotted line shows the median probability of modifying the ad-call-site code.} 
	\label{fig:dis_updating_ad_call_site_code}
\end{figure}
Figure~\ref{fig:dis_updating_ad_call_site_code} shows the probability of modifying ad-call-site
code for every integration strategy. The probability of modifying the ad-call-site code for ad-displaying apps that use the mixed or the scattered strategies is well above the median.
We also observe that the median probability of modifying the ad-call-site code for apps that use the mixed strategy is twice that of the median probability of modifying ad-call-site code for apps that use the external-mediation strategy.
This result indicates one of the drawbacks of using the scattered strategy since app developers would need to modify the ad-call-site code in a much larger proportions of the deployed updates of their apps.

To study whether the modifiability of ad-call-site code is significantly different across the identified four ad library integration strategies, we use the ``Kruskal Wallis test''~\cite{vargha1998kruskal} for the four categorical variables (i.e., the integration strategies) and one metric (the probability of modifying ad-call-site code). We observe that the generated p-value of the test is less than 0.05 indicating that the the modifiability of ad-call-site code is significantly different across the identified four strategies.

\textbf{The proportion of the modified ad-call-site code (\# of ad-call-site code that is modified / \# of total ad-call-site code) is highest in the apps that use the mixed strategy. We observe that app developers mostly modify ad-call-site code when they update their integrated ad libraries.}
Table~\ref{tab:modify_ad_call_site} shows the proportion of the modified ad call-site-code in two cases: when the integrated ad library is updated and when the integrated ad library is not updated.
The proportion (median) of the modified ad-call-site code is zero for each integration strategy (when the ad library is not updated) indicating that app developers usually do not optimize or change their ad library integration code. 
We also observe that in the case when an ad library is updated, the proportion of the modified ad-call-site code for the apps that use the mixed strategy is the highest whereas the proportion is almost zero for the apps that use the external-mediation strategy.
This result is another indication that the mixed strategy may require more effort for maintaining the ad-call-site code over time.

\begin{table}[!t]
	\centering
	\caption{The proportion of modifying ad call-site-code when an ad library is updated and when an ad library is not updated.}
	\label{tab:modify_ad_call_site}
	\resizebox{\columnwidth}{!}{
		\begin{tabular}{lrr}
			\toprule
			\multirow{3}{*}{} & \multicolumn{2}{c}{\% of the modified ad-call-site code (median)}\\ \cline{2-3}
			             \multicolumn{1}{c}{Ad library integration}    & \multicolumn{1}{c}{when the integrated } & \multicolumn{1}{c}{when the integrated} \\ 
			           \multicolumn{1}{c} {strategy}  & ad library is updated & ad library is not updated\\ \midrule
			           Mixed strategy       &   12.5    & 0.0\\
			           Scattered strategy   &   8.0     & 0.0\\
			           Self-mediation strategy  & 3.3   & 0.0\\
			           External-mediation strategy  & 0.0 & 0.0\\

			\bottomrule     
		\end{tabular}
	}
\end{table}

\subsection{The ratio of adding or removing ad libraries}
To understand which ad integration strategy is more flexible for modifying (adding or removing) ad libraries, we calculate \textit{the ratio of adding/removing ad libraries} for each integration strategy. \textit{The ratio of adding/removing ad libraries} is the ratio of the number of updates of an app in which the app developer adds or removes an ad library to the total number updates of the app.

\begin{figure}[!t]
	\centering
	\includegraphics[width=0.45\textwidth]{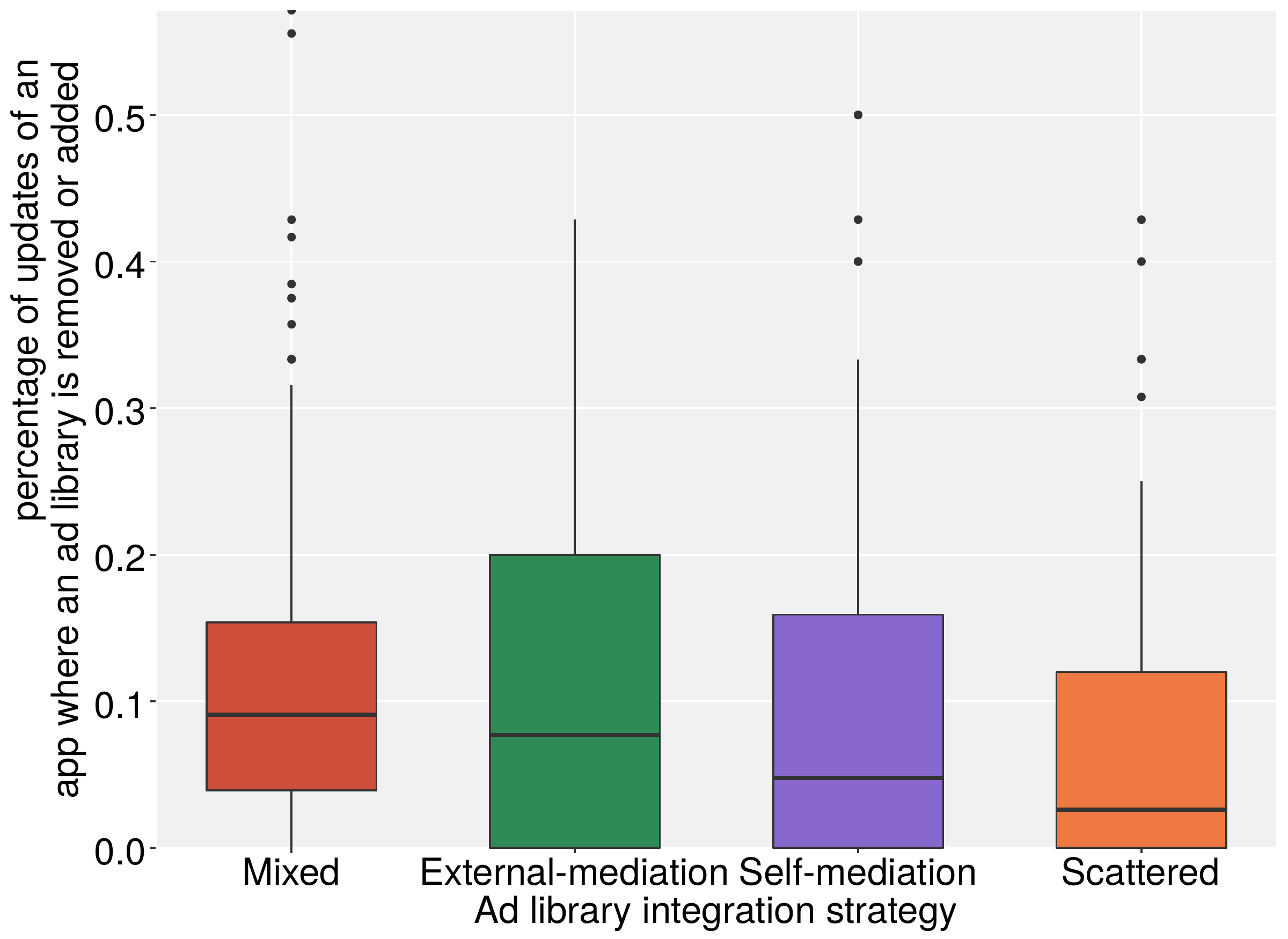}
	\caption{The ratio of adding/removing ad libraries for each of the integration strategies.} 
	\label{fig:flexibility_ad_serving_apps}
\end{figure}

\textbf{The mixed strategy provides app developers with the highest flexibility.} Figure~\ref{fig:flexibility_ad_serving_apps} shows the \textit{the ratio of adding/removing ad libraries} for each identified strategy. The mixed strategy has the highest ratio value. 
We identify ad libraries that are added or removed in the cases of mixed strategy and observe that all these ad libraries are supported by the currently available external-ad-mediators that are currently in use by these apps.
One explanation of this result is that developers do not need to write or update any code to add or remove ad libraries.
This hypothesis explains as well the high \textit{ratio of adding or removing ad libraries} for the external-mediation strategy.

\section{Implications}
\label{sec:Implications_And_Discussions}
In this section, we describe the implications of our study of ad library integration practices for ad library developers.

\textbf{The developers of the Google AdMob should spin out their functionality for uniquely identifying a user's device out of their ad library.}
As described in Section~\ref{sec:RQ1_Results}, analytics libraries have a dependency on the Google AdMob ad library.
These analytics libraries depend on one of the packages of the Google AdMob ad library named ``com.google.android.gms.identifier'' for the unique identification of a user's device.
These analytics libraries need this functionality to track a user's in-app behavior.
However, the main purpose of an ad library is to serve ads.
This unusual dependency on the Google AdMob increases the size of many apps that use these analytical libraries.
Hence, developers of the Google AdMob ad library should rethink their design and offer a separate library for uniquely identifying a user's device.

\textbf{Ad library developers should improve their external-ad-mediators by (1) enabling the integration of new ad libraries at run-time and (2) increasing the supported ad libraries.}
In Section~\ref{sec:Discussions}, we observed that app developers use the mixed strategy to achieve the highest flexibility (i.e., continuously adding or removing an add library). 
To improve the flexibility of the external-ad-mediators, ad library developers need to provide some standardized interfaces to enable the integration of new ad libraries for app developers at run-time instead of only at design time.
For example, the Google AdMob ad library has started to offer an SDK-less mediation feature that enables app developers to add or delete any new ad library by re-configuring their Google ads account (without a need for deploying an update that adds/removes the required ad libraries)~\cite{google_sdk_less}.

Another dimension for improving the external-ad-mediators  is to add support for a large number of ad libraries. For example, in Section~\ref{sec:RQ2_Results}, we observed that 73.9\% of the ad-displaying apps that use the mixed integration strategy call at least one ad library that is not supported by any external-ad-mediators.
Hence, we recommend ad library developers who offer external-ad-mediators to support more ad libraries so that app developers can choose different ad libraries and maximize ad revenue.

\textbf{Ad library developers should offer more feature control over the selection of ad libraries.}
In~Section~\ref{sec:RQ2_Results}, we observed that the external-ad-mediator selects an ad library from the integrated ad libraries based on dynamically calculated eCPM value. 
Although this selection process is useful for accurately estimating the revenue for a served ad, it might prevent an app from achieving an improved user-engagement as the process is not customizable.
In addition, this process is not transparent to app developers (as eCPM is calculated dynamically) and does not allow app developers to control the selection of ad libraries.
We observe that app developers write code (representing a median of 8\% of the total number of classes of the app code) for their self-mediator which offers them custom control when selecting ad libraries based on a preferred list of ad libraries (e.g., a list of ad libraries based on the popularity of ad libraries in a country) or custom app events.
Therefore, we recommend ad library developers to offer a more configurable interface for their external-ad-mediators so that app developers can have more control in selecting ad libraries when they desire.

\section{Threats to validity}
\label{sec:Limitations_And_Threats}
\noindent\textbf{Construct validity:}
App developers can obfuscate their code using obfuscation tools (e.g., Proguard) before releasing their apps. Although Google Admob requires app developers to keep the names of the ad-show methods non-obfuscated (so apps do not get issues while displaying ads)~\cite{Google_ads_dev_proguard}, we cannot assure that all the studied ad libraries mandate the same requirement. Hence, code obfuscation can impact our results as follows. First, if developers obfuscated their package names, our approach will not be able to identify whether an app contains an ad library (as our approach depends on the library package name to identify apps that integrate ad libraries). Moreover, if developers obfuscated the method name of their code, our approach cannot identify whether an app displays ads (as we depend on the usage of the ad-show methods). 
To assess the impact of code obfuscation on our analysis, we measured the percentage of classes, methods and packages that are obfuscated in our studied apps (i.e., 1,837). 
We followed a similar approach to identify the code obfuscation that is employed by Li 
et. al.~\cite{Li_Common_Library}.  
We find at least one obfuscated package and method in 18\% and 39\% of the studied apps, respectively. 
However, in our further investigation on how much code is obfuscated in an app, we find that only 0.5\% (median) of the methods in an app are obfuscated. 
This result shows 
that there can be an impact (albeit a small percentage) that this 0.5\% of the obfuscated methods include show-ad methods of ad libraries.

To identify the apps that display ads, our approach depends on identifying the calls to the show-ad methods of ad libraries. Hence, if an app is obfuscated, our approach would miss identifying such apps as ad-displaying apps. As shown in Table~\ref{tab:stats_app_serve_ads}, 530 apps do not contain any of the identified ad library packages. Out of these 530 apps, 225 apps (12\% of our overall studied 1,837 apps) contain obfuscated methods in their app code. Since we are not the owner of such 225 apps, we cannot assure whether they integrate ad libraries for displaying ads. Hence, our approach may miss identifying ad-displaying apps for these 225 apps. To overcome the issue of missing the identification of ad-displaying apps, we extensively studied a large number of apps that display ads to understand how app developers integrate ad libraries into their apps on a large scale. It should be noted that our objective of this study is to understand the approaches that app developers use to integrate multiple ad libraries instead of providing an approach on how to reverse engineer obfuscated methods. Future studies could extend our work by including more apps from different app stores.

Identifying the ad libraries among the many integrated third-party libraries is a non-trivial task. 
First, we identify packages using the regular expression [aA][dD] (following a similar approach as proposed by Ruiz et al.~\cite{updates_ad_library}). 
Then, we manually search online each identified package to determine whether it corresponds to an ad library. 
However, there is a chance, albeit an extremely rare one, that an ad library exists for which we cannot find any web reference.
To measure the comprehensiveness of our identified integration strategies for ad libraries, we randomly examined a statistical representative sample of 65 apps out of the 1,076 studied ad-displaying apps. 
Such a sample would provide us with a confidence level of 90\% and a confidence interval of 10\%. 
To eliminate any bias in our evaluation results, we do not include the 62 apps that we initially used for identifying the integration strategies for ad libraries. 
For each of the examined apps, we manually examine the app code (using the Understand tool~\cite{understand_tool}) and validate the integration strategies for ad libraries. 
We observe that our proposed rules correctly identify the integration strategies for ad libraries in each of the studied 65 apps. 

Our approach for detecting ad libraries could suffer from both false positives and false negatives. 
In the case of false positives, an app can have dead code with the word “ad” in its class name or a popular third-party library might include ads mischievously. 
However, we studied top apps where we feel the chances of these concerns to occur are extremely low. 
The chances of an app to not remove dead code (and associated ad libraries) is quite low since mobile apps are very mindful of the size of their binaries. 
The chances of an app being malicious are very low again -- recall we are looking at the top apps in the market. 
In the case of false negatives, there might be many ad networks that do not have the term ``ad'' in its APIs. 
The chances of an ad library not having the word ``ad'' is in it is quite low. We also retrieved a list of ad libraries that is curated by AppBrain~\cite{appbrain_ad_networks} and find that the concern is not valid for that curated list. The list contains 120 ad libraries. 
Of these 120 ad libraries, we identified 63 ad libraries that are integrated by the studied apps. 
For the remaining 57 ad libraries, we read their documentation and GitHub packages and validate that their API contains ``ad'' keyword.

To identify the display ad methods, we read the documentation of the studied ad libraries and summarized the list of methods that are currently used for displaying ads. 
However, APIs may evolve, and developers of ad libraries might rename such methods (display ad methods). 
To determine whether such show-ad methods were changed during the evolution of their libraries, we investigated all the historical versions that are released during our study period of the top ten popular ad libraries as follows. 
First, we identified the methods that are currently used for displaying ads from the documentation of the selected ad libraries. 
Then, for each version of the selected ad library, we examined whether the identified methods exist in the prior versions.
We observed that all identified methods exist in the prior versions. We did observe that a few ad libraries (e.g., Unity Ads) added new parameters to these methods. 
However, they did not change the name of these methods. For example, developers of Unity Ads did not change the name of the show method in version 2.0.0 but changed the parameter of the show method (i.e., developers changed \textit{``public void show (Map \textless String, Object\textgreater ~options)''} to \textit{``public void show (activity,  placementId)''} in version 2.0.0). 

One possible threat in our analysis on the modifiability of ad-call-site code is code obfuscation of the studied apps as we cannot track the changes in code statements of obfuscated code. 
Hence, in this analysis, we filtered the obfuscated classes by following the approach of Li et al. ~\cite{Li_Common_Library} and studied the classes of the app code that are non-obfuscated. 
As illustrated in our discussion about code obfuscation, the percentage of obfuscated code is only 0.5\% (median). 
Consequently, there might be cases (albeit a small percentage) that the obfuscated code has modifications to the ads-call cite statements.

Since native code in Android apps is deployed in the app as executable and linkable format (ELF) files~\cite{wiki_elf}, using static analysis tools cannot generate all the dependency links accurately~\cite{native_ndss}. 
We observe that only 6.5\% of the studied apps (121 apps out of 1,837) use native code in our dataset. 
We identified 69 of them integrate ad libraries (i.e., 6\% of the studied ad-displaying apps). 
Studying native apps using static analysis tools is difficult and could introduce false-positive cases in our analysis of the integration strategies for ad libraries. 
Hence, we removed these 69 apps from our analysis of the integration strategies for ad libraries. 
Since these apps are only 3.8\% of the studied apps (69 apps out of 1,837), we believe that they will not drastically impact our overall study of the integration strategies for ad libraries.

\noindent\textbf{External validity:}
In this study, we only focused on the top free-to-download Android apps from the Google Play Store as these apps have a large user-base. 
Hence, these apps are likely to follow the in-app advertising model to earn revenue. 
Future studies should broaden the scope of our study and investigate how our findings apply to ad libraries that are integrated into other types of apps, such as non-free apps, Windows apps, or iOS apps.

According to the documentation of the top ten most used ad libraries in our dataset, the standard way of displaying ads (e.g., full-screen ads) is to call display methods (e.g., showAd()) of the integrated ad libraries~\cite{AdMob_GetStarted,MoPub_GetStarted} from an activity. 
However, we do not claim that this is the only way to display ads. 
For example, app developers can display ads using intents or background services\footnote{https://stackoverflow.com/questions/14313641/show-admob-ads-from-service-context-android}, which is not a recommended practice for displaying ads~\cite{Admob_back_thread}. 
In addition, malicious apps and adware apps can also display ads using background threads, which can be difficult to detect using static analysis mechanisms. 
However, we analyzed popular apps that are actively maintained and present in the Google Play Store, and they are less likely to be malicious or adware apps. 
Additionally, our scope for this research is to study how popular apps integrate ad libraries using standard practices.

In Section~\ref{sec:Implications_And_Discussions}, we point out three implications for ad library developers based on our study so that ad library developers could improve their ad libraries. 
We cannot deny that there might be some business constraints for improving ad libraries. 
For example, supporting the integration of all ad libraries in the ad market is not a simple task and might be influenced by competitive business logic instead of technical challenges. 
However, our findings and suggestions show the current needs and the potential directions for improving the design of third-party ad libraries. 
Hence, our study could be helpful for ad library developers who wish to improve their ad libraries.

\noindent\textbf{Internal validity:}
In our analysis for identifying strategies for integrating multiple ad libraries, we manually investigated apps that integrate more than one ad library.
In this analysis, we cannot deny the possibility of misinterpreting the identified strategies for integrating multiple ad libraries since we are not the original developers of the studied apps. 
To mitigate this threat, the first and the second author leveraged the Understand tool to analyze the call graph of the apps, carefully investigated each of the sampled apps, and consolidated their results.

App users are prone to repackaged or piggybacked apps. 
Hence, app stores (e.g., the Google Play Store) continuously remove piggybacked apps or malware apps~\cite{Google_Removes_Malware_Apps,Li_apps_remove_msr_18}. 
In our study, we selected 1,837 top free-to-download apps in the Google Play Store. 
These apps have a large user-base, and they are popular in the Google Play Store for multiple years. 
Therefore, our studied apps are less likely to be malicious apps.

There are different possibilities that can drive app developers to group their code in a single component. 
For example, poorly designed apps can be one of the reasons for this behaviour (e.g., app developers just group all their code in a single package). 
However, based on our definition of the self-meditation strategy, an app contains a self-mediator component whenever: (1) there is a centralized component that handles all the calls to ad libraries and (2) this component is separate from the main activity code components. 
Our definition means that the design of the self-meditation strategy encapsulates ad handling mechanism in a single package. 
However, the identified self-mediation components can be poorly designed if they contain non-ad related features. 
Assessing the quality of mobile apps architecture is out of the scope of this work. Further work can analyze the architectural quality of mobile apps.

To validate our approach for identifying the ``self-mediation'' strategy, we manually investigate a randomly selected sample of 20 apps that use the ``self-mediation'' strategy. We observe that all of these 20 apps have a centralized package that is mainly responsible for managing the integration of multiple ad libraries. 
For example, the ``Agar.IO'' app (a popular app in the Game category) has the package ``com.miniclip.ads'', which manages the integration of 12 ad libraries.

\section{Related work}
\label{sec:Related_Work}
Prior research mainly studies the updates of ad libraries, the cost of ad libraries and the security and privacy issues surrounding ad libraries. 
Our study is the first to investigate the integration practices of ad libraries.
We briefly highlight the related works as follows:

\subsection{The updates of ad libraries}

Ruiz et al.~\cite{updates_ad_library} analyzed 120,981 free-to-download apps in the Google Play Store and conducted an empirical analysis on the rational for updating ad libraries.
The authors observed that ad libraries are updated frequently in 48\% of the studied apps. 
They also observed that updating the interaction between ads and app users, integrating new types of ads, bugs related to memory management, and the improvement of the privacy of collected personal information are the main reasons for updating ad libraries.

Salza et al.~\cite{Salza:2018:DUT:3196321.3196341} conducted an empirical study on the evolution history of 291 apps from the F-Droid repository to study how mobile app developers perform updates of third-party libraries including ad libraries.
The authors observed that developers usually upgrade towards a newer version. 

Derr~\cite{keep_me_updated} et al. studied the updatability of third-party libraries (including ad libraries). 
The authors  analyzed the updatability of the integrated libraries of 1,264,118 apps from the Google Play Store. They observed that in 85.6\% of the cases the integrated ad libraries can be updated to at least one version without any code changes.

\subsection{The cost of ad libraries}
Ruiz et al.~\cite{israel_app_rating_ads_library} analyzed 236,245 apps of the Google Play Store to study the effect of ad libraries on the rating. 
The authors observed that there is no relation between the number of integrated ad libraries and the rating of an app. However, they observed that integrating certain ad libraries could negatively impact the rating of an app.

Gui et al.~\cite{ads_hidden_cost}  analyzed 21 apps from the Google Play Store to study five types of costs due to the integration of ad libraries: performance, energy consumption, network, maintenance of ad-related code, and user reviews.
The authors observed that the cost of ads in terms of performance, energy and bandwidth are the most concerning. 
They also observed that complaints related to ads had a negative impact on the rating of an app.

Gao et al.~\cite{Gao_intelli_ad} 
designed a tool named IntelliAd to automatically measure the ads-related consumption (e.g., memory) on apps. 
In another study, Gao et al.~\cite{Gao_ad_schema_cost} used the IntelliAd tool and analyzed 12 ad schemes that are used in 104 Android apps to measure and compare the performance cost of different ad schemes. 
They observed that ad schemes are significantly different and recommend app developers to choose appropriate ad providers and ad sizes.

To study the ad network traffic, Vallina-Rodriguez et al.~\cite{Vallina-Rodriguez_ads_breaking} analysed the dataset of a European mobile career with more than 3 million subscribers. The author observed that ads account for 1\% of all mobile traffic in the data and, static images and text files are likely to be re-downloaded. 
To limit the energy and network signalling overhead, the authors built a prototype implementation using the caching mechanism which shows an improvement of 50\% in energy consumption for offline ad-sponsored apps. 

Mohan et al.~\cite{Mohan_prefetching} studied the communication costs for serving ads by analyzing 15 Windows phone. The authors observed that ad modules consume a significant part of an app's energy and the overhead of ads is bigger in apps with no or small network activity.
To reduce the energy of an app that displays ads, the authors proposed an solution of prefetching ads. The authors analyzed the logs of 1,693 Windows phone users over one month and 25 iPhone users over one year to predict app usage from historical data and built time-based models to predict available ad slots in future. The entropy-based evaluation result of their approach shows that the approach is capable to reduce energy consumption of client devices by 50\%.

Li et al.~\cite{Li_Common_Library} analyzed 1.5 million apps that use 1,113 third-party libraries
and 240 ad libraries to investigate the use of commonly integrated libraries. 
Their study showed that the most used library is the Google ad library (AdMob). 
Li et al. also found that a significant portion of apps that used ad libraries are apps that are flagged by virus scanners.
In our study, we focus on top rated apps to avoid dealing with malicious and spam apps.

\subsection{The security of ad libraries}
Book et al.~\cite{longitudinal_Permission_Analysis} studied the evolution of the requested permission of ad libraries. 
The authors analyzed the integrated ad libraries of 114,000 apps for this study. They observed that the use of permission has increased over time and most of the requested permissions of ad libraries are risky in terms of the privacy and security of app users.

Kim et al.~\cite{kim} analyzed the protective security measures of the four ad libraries (Google AdMob, MoPub, AirPush, and AdMarvel) against malicious advertising.
The authors observed that malicious ads can collect sensitive information about a user with the help of permissions such \textit{WRITE\_EXTERNAL\_STORAGE} and \textit{READ\_EXTERNAL\_STORAGE} which give access to external storage.  

Dong  et  al.~\cite{Dong_ad_fraud}  studied  ad  fraud (e.g., cheating advertisers with fake ad clicks) in mobile apps and proposed an automated approach for detecting ad fraud. 
Their approach achieves 92\% recall and 93\% precision on a manually validated data set of 100 apps. A further study on 12,000 ad-supported apps that use 20 unique ad libraries showed that no ad library was exempt from fraudulent behaviours and AppBrain ad library is the most targeted ad library for ad fraud.

\section{Conclusions}
\label{sec:Conclusion}
In the mobile app economy, ad libraries play an integral role.
App developers integrate one or many ad libraries to display ads and gain revenue based on user interactions with the displayed ads.
Even though ad libraries play an essential role in the app ecosystem, prior studies have not explored how these libraries are integrated by app developers. 
In this paper, we analyze 35,459 updates of the 1,837 top free-to-download apps to study the ad library integration practices.
Our findings highlight how app developers leverage several ad libraries to display ads. Moreover, we noted several limitations of current ad libraries and offered suggestions for ad library developers to better serve the needs of app developers.

\balance{
\bibliographystyle{abbrv}
\bibliography{Ads_References}
}
\begin{IEEEbiography}[{\includegraphics[width=1in,height=1.25in,clip,keepaspectratio]{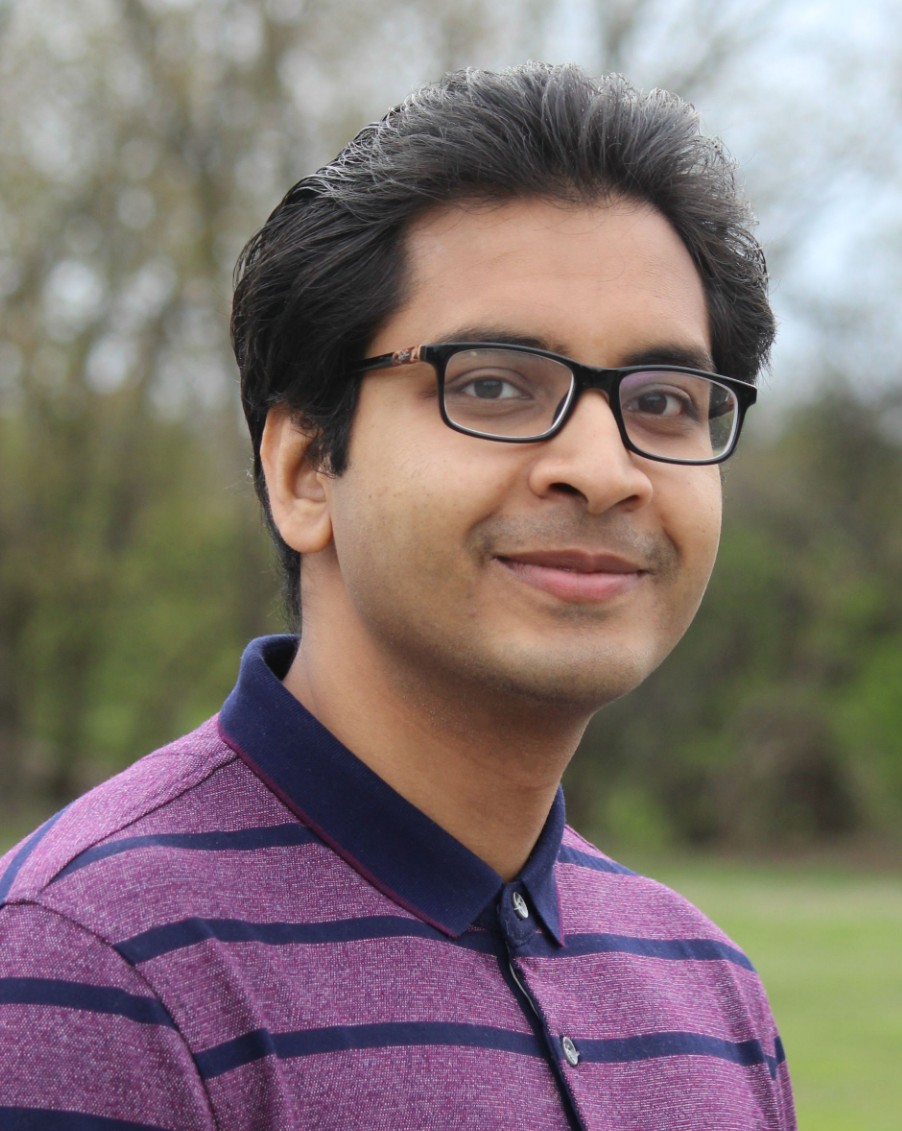}}]{Md Ahasanuzzaman} is  a graduate student in the Software Analysis and Intelligence Lab (SAIL) at Queen's University, Canada. His research interests include mining software repositories, analyzing mobile apps and app stores, analyzing community question answering sties, natural language processing, and machine learning. His works got published in top venues of Software Engineering (e.g., MSR, SANER, and EMSE). He obtained his BSc from the University of Dhaka (Department of Computer Science and Engineering), Bangladesh. He has been awarded prestigious  awards,  such  as  Dean’s  scholarship  award  and  Prime Minister Gold Medal for his outstanding achievements in the B.Sc program. More about Md Ahasanuzzaman can be read on his website:https://ahasanuzzaman.com/research/
\end{IEEEbiography}
\begin{IEEEbiography}[{\includegraphics[width=1in,height=1.25in,clip,keepaspectratio]{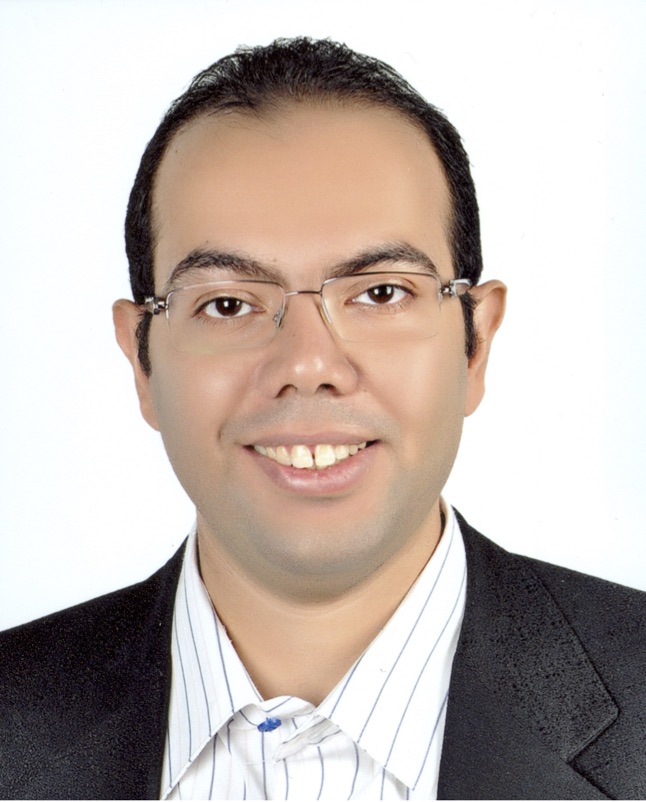}}]{Safwat Hassan} currently works as a Postdoctoral Fellow in the Software Analysis and Intelligence Lab (SAIL) at Queen's University, Canada.
Hassan worked as a software engineer for ten years in different corporations, including the Egyptian Space Agency (ESA), HP, EDS, VF Germany (outsourced by HP), and Etisalat. During his ten years in the software industry, he worked on a variety of software systems, such as web-based systems and embedded systems. He also participated in diverse project roles (e.g., design service, customer support, and R\&D) across multiple business domains (e.g., telecommunication, supply-chain, and aerospace). His research interests include data mining for software engineering, mobile app store analytics, software architecture, system anomaly prediction, continuous integration, and software performance analytics.
More about Safwat Hassan can be read on his website: https://safwathassan.com
\end{IEEEbiography}
\begin{IEEEbiography}[{\includegraphics[width=1in,height=1.25in,clip,keepaspectratio]{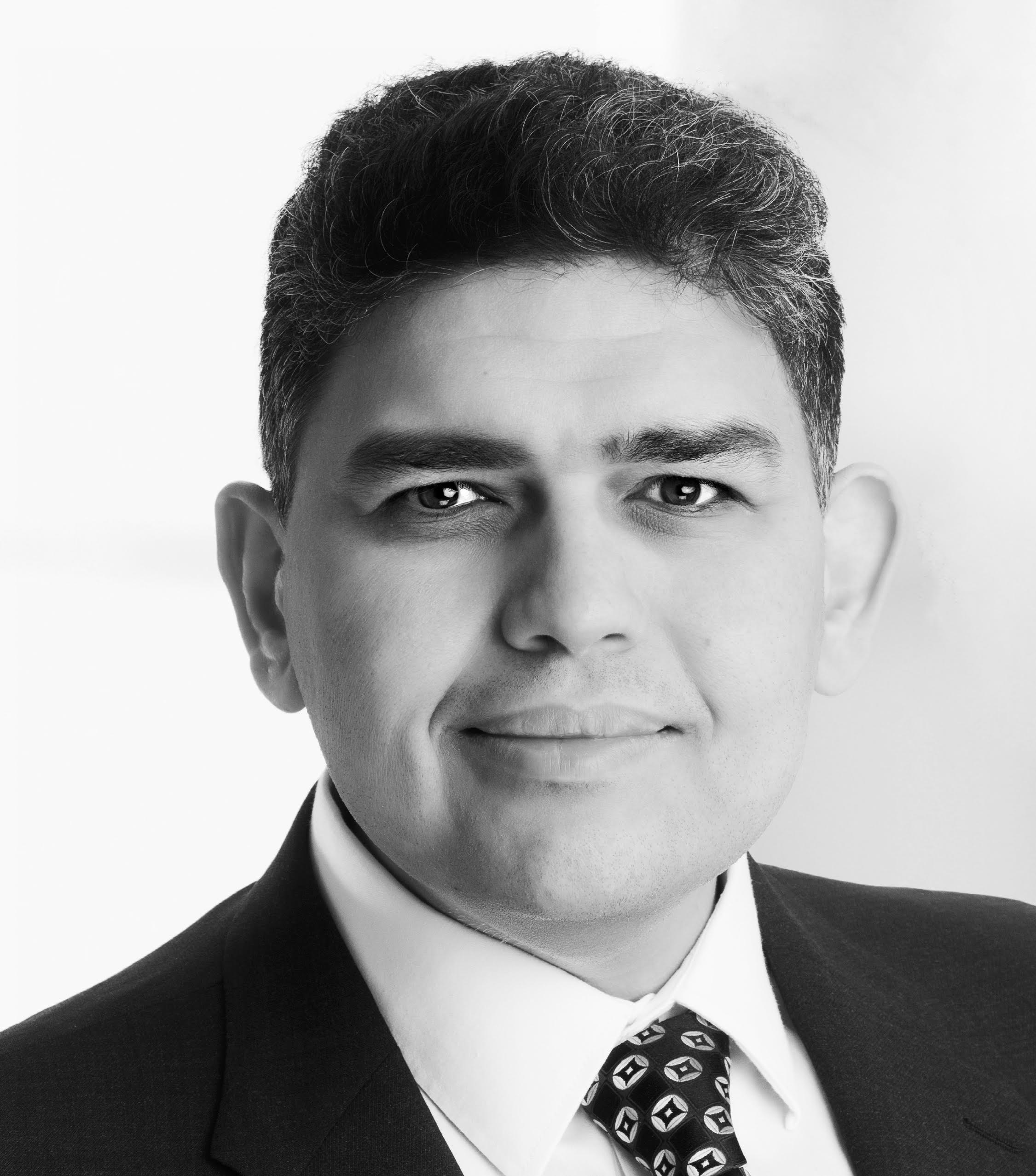}}]{Ahmed E. Hassan}  is an IEEE Fellow, an ACM SIGSOFT Influential
Educator, an NSERC Steacie Fellow, the Canada Research Chair (CRC) in Software Analytics, and the NSERC/BlackBerry Software Engineering
Chair at the School of Computing at Queen's University, Canada. His
research interests include mining software repositories, empirical
software engineering, load testing, and log mining. He received a PhD in Computer Science from the University of Waterloo. He spearheaded the creation of the Mining Software Repositories (MSR) conference and its research community. He also serves/d on the editorial boards of IEEE Transactions on Software Engineering, Springer Journal of Empirical Software Engineering, and PeerJ Computer Science. More information at: http://sail.cs.queensu.ca/
\end{IEEEbiography}

\end{document}